\documentclass[useAMS,usenatbib]{mn2e}
\usepackage[dvips]{graphicx}
\usepackage{longtable}

\title[Reflections on Reflexions]{Reflections on Reflexions: 
	I. Light Echoes in Type Ia Supernovae}
\author[Ferdinando Patat]
  {F.~Patat$^1$\\
  $^1$European Southern Observatory, Karl-Schwarzschild str. 2, 
  D-85748 Garching bei M\"unchen, Germany.\\}

\begin{document}

\date{Accepted.......;  Received .......}

\pagerange{\pageref{firstpage}--\pageref{lastpage}} \pubyear{2002}

\maketitle

\label{firstpage}

\begin{abstract}
In the last ten years, observational evidences about a possible
connection between Type Ia Supernovae (SNe) properties and the
environment where they explode have been steadily growing.  In this
paper I discuss, from a theoretical point of view but with an
observer's perspective, the usage of light echoes (LEs) to probe the
CSM around SNe of Type Ia since, in principle, they give us a unique
opportunity of getting a three-dimensional description of the SN
environment. In turn, this can be used to check the often suggested
association of some Ia's with dusty/star forming regions, which would
point to a young population for the progenitors.  After giving a brief
introduction to the LE phenomenon in single scattering approximation,
I derive analytical and numerical solutions for the optical light and
colour curves for a few simple dust geometries.  A fully 3D multiple
scattering treatment has also been implemented in a Monte Carlo code,
which I have used to investigate the effects of multiple
scattering. In particular, I have explored in detail the LE colour
dependency from time and dust distribution, since this is a promising
tool to determine the dust density and derive the effective presence
of multiple scattering from the observed properties.  Finally, again
by means of Monte Carlo simulations, I have studied the effects of
multiple scattering on the LE linear polarization, analyzing the
dependencies from the dust parameters and geometry.  Both the
analytical formalism and MC codes described in this paper can be used
for any LE for which the light curve of the central source is known.
\end{abstract}


%

\section{Introduction}
\label{sec:intro}

In the last few years the study of light echoes (hereafter LEs) in
Supernovae (SNe) has become rather fashionable, since they provide a
potential tool to perform a detailed tomography of the SN environment
(see \citealt{sugerman03} and references therein)
and, in turn, can give important insights into the progenitor's
nature, a matter which is still under debate. Due to the typical
number density of dust particles which are responsible for the light
scattering, LEs are expected to have an integrated brightness
about ten magnitudes fainter than the SN at maximum. For this reason,
a SN Ia in the Virgo cluster is supposed to produce, if any, an echo
at a magnitude V$\sim$21.0. This has the simple consequence that it is
much easier to observe such a phenomenon in a Ia than in any other SN
type, due to its high intrinsic luminosity. As a matter of fact, only
four cases are known: the SNe Ia 1991T (\citealt{schmidt},
\citealt{sparks99}) and 1998bu (\citealt{capp}), and the type 
II SN1987A (\citealt{crotts89}, \citealt{crotts91}, \citealt{xu94},
\citealt{crotts95}, \citealt{xu95}) and 1993J (\citealt{sugerman}). 
As expected, the LE detections for the two core-collapse events
occurred in nearby galaxies: LMC (d=50 Kpc) and M81 (d=3.6 Mpc)
respectively. Now, while a dusty environment is expected for
explosions arising in massive and short-lived progenitors, this is in
principle not the case for the kind of stars which are commonly
supposed to generate Ia events, i.e. old population and low mass. This
is why the detection of substantial dust in the immediate surroundings
of thermonuclear SNe, or at least in a fraction of them, would
indicate a different scenario.

Actually, several authors have pointed out that the observed
characteristics of Ia's, such as intrinsic luminosity, colour, decline
rate, expansion velocity and so on, appear to be related to the
morphological type of the host galaxy (\citealt{flip89},
\citealt{branchvdb}, \citealt{vdb};
\citealt{hamuy96}, \citealt{hamuy00}; \citealt{howell}). Since these objects 
represent a fundamental tool in Cosmology (see for instance
\citealt{bruno}), it is clear that a full understanding of the underlying
physics is mandatory in order to exclude possible biases in the
determination of cosmological parameters like $\Omega$ and
$\Lambda$. In this framework, disentangling between possible Ia
sub-classes is a fundamental step.  In this respect, an important year
in the SN history is 1991, when two extreme objects were discovered,
i.e.  SN1991T (\citealt{flip92a}) and SN1991bg (\citealt{flip92b}). The former 
was an intrinsically blue, slow declining and spectroscopically
peculiar event, while the latter was intrinsically red, fast declining
and also showing some spectral peculiarities.  From that time on,
several other objects sharing the characteristics of one or the other
event were discovered, indicating that these deviations from the
standard Ia were, after all, not so rare. Of course, one of the most
important issues which was generated by the discovery of such theme
variations concerned the explosion mechanism and, in turn, the progenitor's
nature.

The growing evidences produced by the observations in the last ten
years have clearly demonstrated that the sub-luminous events
(1991bg-like) are preferentially found in early type galaxies (E/S0),
while the super-luminous ones (1991T-like) tend to occur in spirals
(Sbc or later). This has an immediate consequence on the progenitors,
in the sense that sub-luminous events appear to arise from an old
population while super-luminous ones would rather occur in
star-forming environments and therefore would be associated with a
younger population. This important topic has been recently discussed
in a work by \citet{howell}, to which I refer the reader for
a more detailed review.  What is important to emphasize here is that
1991T-like events tend to be associated with young environments and
are, therefore, the most promising candidates for the study of
LEs. Or, in turn, if LEs are detected around such kind of SNe, this
would strengthen their association with sites of relatively recent
star formation.

The first case of LE in a Ia (1991T) seems to confirm this hypothesis,
in the sense that the SN was over-luminous and the host galaxy
(NGC4527) is an Sbc and also a liner. Slightly less convincing is the
other known case (1998bu), since the galaxy (NGC3368) is both an Sbc
and a liner, but the SN is not spectroscopically peculiar. The only
characteristic in common with SN1991T is its decline rate $\Delta
m_{15}$, which is lower than average, even though not so extreme as in
the case of 1991T. Nevertheless, the HST observations by
\cite{garnavich}) show that a significant amount of dust must be
present within 10pc from the SN.  Of course no statistically
significant conclusion can be drawn from such a small sample, which
definitely needs to be enlarged. For this reason, during the past
years, I have been looking for new cases, the most promising of which
was represented by SN1998es in NGC~632. This SN, in fact, was
classified as a 1991T-like by \citet{jha}, who also noticed that the
parent galaxy was an S0, hosting a nuclear starburst
\citep{pogge}. Moreover, the SN was found to be projected very close
to a star forming region and to be affected by a strong reddening,
which all together made SN~1998es a very good candidate for a LE
study.

While the present paper is devoted to the theoretical aspects of the
LE phenomenon, in a forthcoming article (Patat et al.  in preparation)
we will be dealing with the application of these results to the
known cases of SNe 1991T, 1998bu and the test case of SN~1998es. In
the same work we will also present high resolution spectroscopy for
the three events and other unpublished data relevant for the LE
discussion.

The paper is organized as follows. After giving a general introduction
to the LE phenomenon in single scattering approximation in
Sec.~\ref{sec:single}, in Sec.~\ref{sec:analyt} I derive analytical
solutions for a thin perpendicular sheet and a spherical shell to
illustrate the effects of forward scattering. In
Sec.~\ref{sec:numeric} I then present a numerical solution for a
double exponential dust distribution, typical of a spiral
galaxy. Single scattering approximation is refined in
Sec.~\ref{sec:ssa}, where I introduce the attenuation and the concept
of LE effective optical depth. The inclusion of multiple
scattering is presented in Sec.~\ref{sec:mc}, where I illustrate the
implementation of a Monte Carlo (MC) code, whose results are shown and
discussed in Sec.~\ref{sec:mcres}. The same code is then used to
compute the LE spectrum (Sec.~\ref{sec:spectrum}) and broad band
$(B-V)$ colour (Sec.~\ref{sec:colour}), which is compared with the
results obtained with the single scattering plus attenuation
approximation.  The effects of a change in the dust mixture are
described and discussed in Sec.~\ref{sec:mixture}. In
Sec.~\ref{sec:polar} I introduce the concepts of MC polarization
calculations, which I have used in a code to study the effects of
multiple scattering, presented in the same Section.  Finally, in
Sec.~\ref{sec:discuss} I give a closing discussion and in
Sec.~\ref{sec:conclusions} I briefly summarize my results.

\section{Single Scattering Approximation}
\label{sec:single}

Starting with the pioneering work by \cite{couderc}, the problem has
been addressed by several authors (see for example \citealt{dwek},
\citealt{chev},  \citealt{schae}, \citealt{emmering}, \citealt{sparks94}, 
\citealt{xu94}, \citealt{sparks97} and \citealt{sugerman03}), who have all 
adopted the single scattering approximation. The only exception is the
work by \citet{chev}, who has used a MC code to study the case of a
spherically symmetric stellar wind with a $r^{-2}$ dust density law.

Given the number of existing publications on this topic, I will give here 
only a brief introduction to the LE phenomenon in the single 
scattering approximation, while I will focus on the analytical and 
numerical solutions which I shall use later on to test the MC code.

\subsection{Integrated echo light curve}
\label{sec:anlc}

\begin{figure}
\centering
\resizebox{\hsize}{!}{\includegraphics{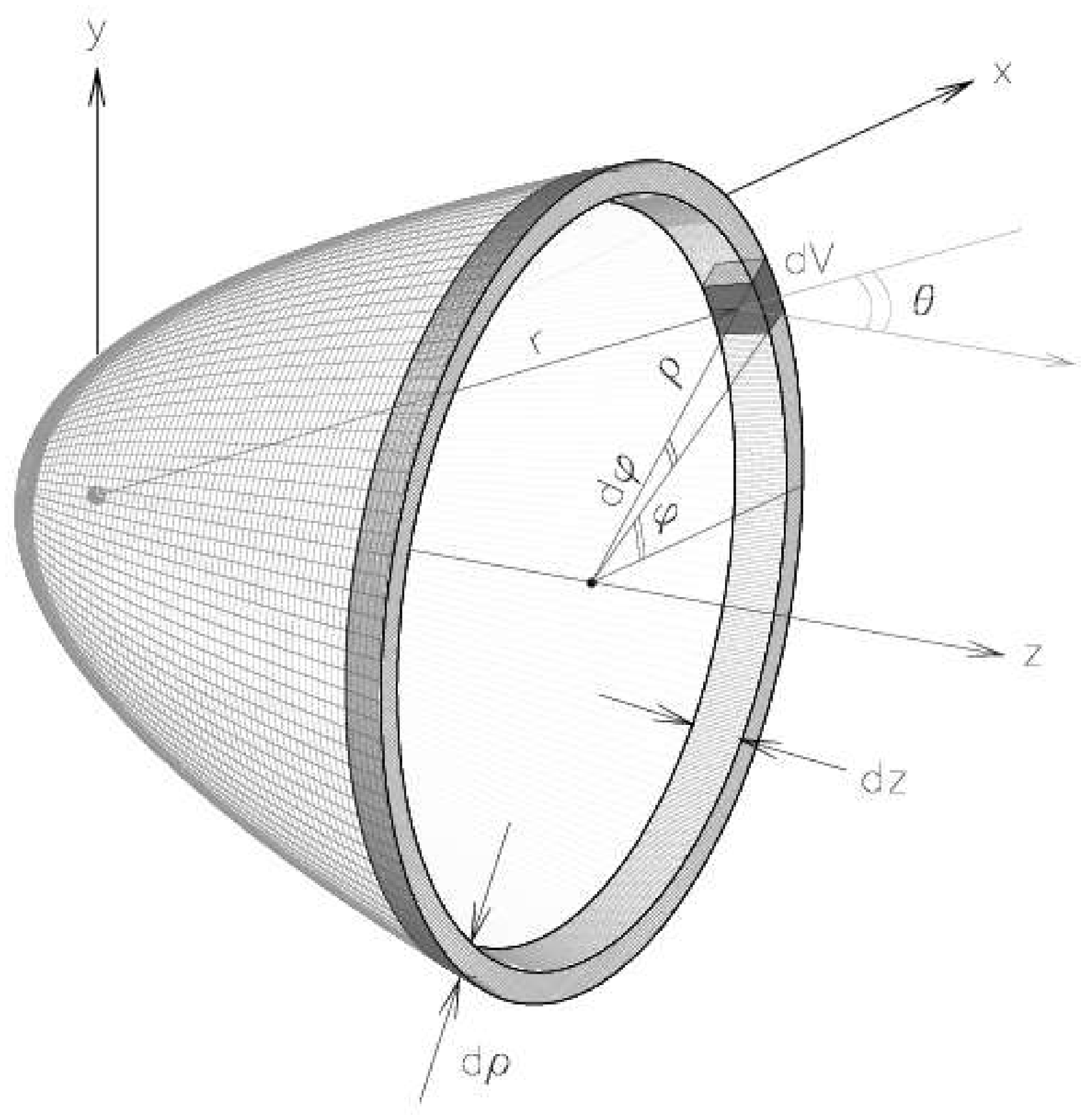}}
\caption{\label{fig:geom} Geometry of the problem in single scattering
approximation.}
\end{figure}

Let us imagine a SN immersed in a dusty medium at a distance $d$ from
the observer. If we then consider the SN event as a radiation flash,
whose duration $\Delta t_{SN}$ is so small that $c\;\Delta t_{SN}\ll
d$, at any given time the LE generated by the SN light scattered into
the line of sight and seen by the observer is confined in an
ellipsoid, whose foci coincide with the observer and the SN
itself. Since $d$ is supposed to be very large compared to all other
geometrical distances, this ellipsoid can be very well approximated by
a paraboloid, with the SN lying on its focus (see for example
\citealt{chev}).  More generally, the paraboloid can be regarded as
the locus of those scattering elements which produce a constant delay
$t$ and for this reason I will refer to it as the iso-delay
surface. If $L(t)$ is the number of photons emitted per unit time by
the source at a given wavelength, the flux of scattered photons per
unit time and unit area which reach the observer at time $t$ is
obtained integrating the flux scattered by the volume element $dV$
(see Fig.~\ref{fig:geom}) over all iso-delay surfaces:

\begin{equation}
\label{eq:F}
4\pi d^2\;F(t)=\int_0^t L(t-t^\prime) \; f(t^\prime) \; dt^\prime
\end{equation}

The $f(t)$ function, which has the dimension of the inverse of a time,
contains all the information about physical and geometrical properties
of the dust and it is defined as follows:

\begin{equation}
\label{eq:f}
f(t) = C_{ext} \; \omega\; \;c
\int_{-ct/2} ^{+\infty} \frac{\Phi(\theta)}{r}
\int_0^{2\pi} n(\varphi,z,t)\; d\varphi \; dz
\end{equation}

where $\varphi$ is defined as in Fig.~\ref{fig:geom}, $r=z+ct$ is the
distance of the volume element from the SN, $n=n(x,y,z)$ is the number
density of scattering particles, $C_{ext}$ is the extinction
cross-section\footnote{It must be noticed that in the literature
$C_{ext}$ is usually given per Hydrogen atom (see for example
\citealt{draine}). In that case, $n$ has to be replaced with the
Hydrogen numerical density $n(H)$.}, $\omega$ is the dust albedo,
$\theta$ is the scattering angle defined by $\cos\theta=z/r$ and
$\Phi(\theta)$ is the scattering phase function, normalized in order
to have $\int_{4\pi}\Phi (\theta )d\Omega=1$.  As usual in this kind
of studies (cfr. \citealt{chev}), I have adopted the formulation
proposed by Henyey \& Greenstein (1941, hereafter H-G), which includes
the degree of forward scattering through the parameter $g$: isotropic
scattering  is obtained for $g$=0 ($\Phi(\theta)=1/4\pi$) while $g$=1
gives a complete forward scattering.  Both empirical estimates and
numerical calculations for the optical wavelength range give $g\approx$0.6
\citep{white}, which corresponds to an average scattering angle
$\overline{\theta}\sim$50$^\circ$. More sophisticated calculations,
including a full Mie's treatment, show that the H-G formulation tends
to underestimate the forward scattering, but the approximation is
reasonably good (see for example \citealt{bianchi}).  As for the dust
albedo, this is usually assumed to be $\omega\approx$0.6
\citep{mathis}.

From Eq.~\ref{eq:F} it is clear that the out-coming scattered signal
is the convolution of the SN signal $L(t)$ with the impulse response
function $f(t)$ and therefore, at any given time $t$, the observer
will receive a combined signal, which contains a sum of photons
emitted by the SN in the whole time range $0-t$. This equation is
formally identical to the one which describes the resulting acoustic
wave reverberation in a given environment (see for example
\citealt{spjut}).

The global flux received by the observer is obtained adding to the
echo contribution the photons coming directly from the SN, after
correcting for possible extinction: 

\begin{equation}
\label{eq:FT}
 F_T(t) = \frac{L(t)}{4\pi d^2} \; e^{-\tau_d} + F(t)
\end{equation}

where

\begin{equation}
\label{eq:taud}
\tau_d=\int_0^{+\infty} C_{ext} \; n(0,0,z) \; dz
\end{equation}

is the dust optical depth along the line of sight. Of course, if other
dust is present but it does not contribute to the LE, because it is
too far from the source, both the SN and the LE would suffer by
additional extinction.

Due to the very short duration of the SN burst and in the single
scattering approximation, there is a very good correspondence between
the observed impact parameter $\rho=\sqrt{ct \; (2z+ct)}$ and the
distance $r$ of the scattering material from the SN (see
Fig.~\ref{fig:geom}). This can be easily calculated as

\begin{equation}
\label{eq:eqdist}
r=\frac{1}{2}\;
\left ( \frac{\rho^2}{ct} + ct
\right )
\end{equation}

and it offers the unique possibility of measuring the third dimension of a
resolved LE image with a spatial resolution of the order of
$c\Delta t_{SN}$, provided that $d$ is known with some approximation.
I note that this is always true, no matter what the dust geometry is,
only provided that the dust optical depth is not too large to cause
multiple scattering, which would introduce some degree of degeneracy
in the $\rho$ vs. $r$ relation (see Sec.~\ref{sec:mcres}).

\subsection{Echo surface brightness}
\label{sec:integx}

If, for a given point $(x,y)$ on the sky plane, we consider the sum of
all contributions along the $z$ axis, we can easily compute the echo
surface brightness $\Sigma$ at any given time as:

\begin{equation}
\label{eq:surfbr}
\Sigma(x,y,t) = \frac{2c \; \omega \; C_{ext}}{4\pi d^2}
\int_0^t \; \frac{L(t-t^\prime)}{c^2 t^2 + \rho^2} \;
n(x,y,t) \; \Phi(\theta) \; dt^\prime
\end{equation}

The explicit dependency of $\Phi$ from $x,y$ and $t$ through
the scattering angle $\theta$ can be obtained remembering that
$\cos\theta=z/(z+ct)$ which, using the paraboloid equation (cfr.
\citealt{chev}), can be
reformulated as:

\begin{equation}
\label{eq:costheta}
\cos\theta(\rho,t)=
\left (\frac{\rho^2}{c^2 t^2}-1\right )
\left (\frac{\rho^2}{c^2 t^2}+1\right )^{-1}
\end{equation}

\section{Simple analytical solutions}
\label{sec:analyt} 

Eqs. \ref{eq:f} and \ref{eq:F} can be solved numerically, using the
observed light curve of a typical Ia for $L(t)$. Nevertheless, 
instructive results can be obtained analytically if one assumes that
the SN light curve is a flash:

\begin{equation}
\label{eq:flash}
L(t) = \left\{ \begin{array}{ll} L_0 & \mbox{for $t \leq \Delta
t_{SN}$} \\ 0 & \mbox{for $t > \Delta t_{SN}$} \end{array} \right.
\end{equation}

In this approximation and for $t\gg \Delta t_{SN}$, the convolution in
Eq.~\ref{eq:F} becomes trivial and gives:

\begin{equation}
\label{eq:Fsimple}
4\pi d^2\; F(t) = L_0\;\Delta t_{SN}\; f(t)
\end{equation}

and hence the whole problem reduces to compute $f(t)$ defined by
Eq.~\ref{eq:f}. The flash duration can be estimated from the observed
light curve as $L_0\;\Delta t_{SN}=\int_0^{+\infty} L(t)\;dt$, and
turns out to be of the order of 0.1 years for a typical Ia. The use of
H-G formulation with $g\neq$0 in Eq.~\ref{eq:f} makes the analytical
integration possible only under certain conditions. For example,
\citet{chev} has found an analytical solution for the case of a
density law $n\propto r^{-2}$, typical of stellar winds. Here I will
focus on two simpler cases only, those of a thin perpendicular sheet
and a thin spherical shell, which are useful to illustrate the effect
of forward scattering. In both geometries the solution becomes very
easy, since $\mu$ can be considered constant along the iso-delay
surfaces where $n\neq$0.

Let me first consider a thin sheet perpendicular to the line of sight,
placed at a distance $R$ from the SN and with a thickness $\Delta
R$. A limit on $\Delta R$ in order to fulfill the above condition can
be derived differentiating the relation between $\mu$ and $z$ (see
Sec.~\ref{sec:single}) and imposing $\Delta \mu\ll 1$. This gives:

\begin{displaymath}
\Delta R \ll \frac{(R+ct)^2}{ct}
\end{displaymath}

Having this in mind and recalling that for $z=R$, one has that
$\mu=z/r\equiv R/(R+ct)$, Eq.~\ref{eq:f} can be
integrated and it gives:

\begin{equation}
\label{eq:thinsheet}
f(t) \simeq
\frac{\omega \;C_{ext} \; c\;n_0\;(1-g^2)}{2(1+g^2-2g\frac{R}{R+ct})^{3/2}}\;
\ln \left( 1+\frac{\Delta R}{R+ct} \right)
\end{equation}

where, with the additional assumption that $\Delta R \ll R+ct$
(not necessarily included in the previous condition), the logarithmic term
can be approximated with $\Delta R/(R+ct)$, as done by
\citet{capp}. As for the integrated luminosity, I notice that under these
circumstances the time dependency is very mild. Therefore, using 
typical values for the relevant parameters ($g$=0.6, $\omega$=0.6, 
$\Delta t_{SN}$=0.1 yrs) and substituting Eq.~\ref{eq:thinsheet} in
Eq.~\ref{eq:Fsimple}, the normalized LE flux can be written as

\begin{displaymath}
4\pi d^2\;\frac{F(t)}{L_0}\approx 0.3\;\frac{\tau_d}{R}
\end{displaymath}

where I have posed $\tau_d=C_{ext}\; n_0\;\Delta R$ (see
Eq.~\ref{eq:taud}). This basically means that, at least for a distant
cloud ($R\gg ct$), the LE luminosity is inversely proportional to $R$
and directly proportional to $\tau_d$.  For example, for a sheet placed
at $R$=100 lyrs from the SN, in order to produce a LE 10 mags fainter than
the SN at maximum, one needs an optical depth $\tau_d\simeq$0.03.

In general, the resulting echo will appear as a ring of radius
$\rho_R=\sqrt{ct(2R+ct)}$ and thickness $\Delta \rho_R\simeq
ct\;\Delta R/\rho_R$, with roughly constant surface brightness.

The overall effect of forward scattering can be evaluated directly
from Eq.~\ref{eq:thinsheet}. In fact, for $R\gg ct$, it is
$R/(R+ct)\simeq$1, so that the forward scattering term becomes
approximately $(1+g)/(1-g)^2$. This means that going from $g=0$ to
$g=0.6$, the echo maximum luminosity increases by more than a factor
of 10.  Moreover, as a consequence of the loss in the scattering
efficiency due to the increase of the scattering angle with time, the
resulting light curves get steeper. This is particularly true when the
dust is not too far from the central source, while for increasing
values of $R$ the time dependency of the forward scattering term tends
to disappear (see Eq.~\ref{eq:thinsheet}).

The inclusion of forward scattering produces also a sensible
flattening of the surface brightness profile. In fact, an increase of
$g$ causes the material which is close to the SN to be less efficient
in scattering the incoming photons, and this tends to balance with the
fact that more distant material receives less photons from the SN but
it is more effective in deflecting its light.  Furthermore, for
homogeneous distributions, the surface brightness at $\rho=ct$
decreases by a factor $(1-g^2)/(1+g^2)^{3/2}$ which, for $g=0.6$,
corresponds to about one magnitude. This is relevant for the detection
of the maximum polarization ring (see \citealt{sparks94} and the
discussion in Sec.~\ref{sec:polar} here).

Another example of analytical solution in the case of $g\neq$0 is the
one of a thin spherical shell of radius R and thickness $\Delta R\ll
R$. In this case, since $R=z+ct$, we have that $\mu=(R-ct)/R$
and therefore the solution of Eq.~\ref{eq:f} gives:

\begin{displaymath}
\label{eq:thinshell}
f(t) \simeq
\frac{\omega\; C_{ext} \; c\;n_0\;(1-g^2)}{2(1+g^2-2g\frac{R-ct}{R})^{3/2}}\;
\ln \left( 1+\frac{\Delta R}{R} \right)
\end{displaymath}

Also in this case the LE would appear as a ring, whose radius
$\rho_R=\sqrt{ct(2R-ct)}$ grows, reaches a maximum at $t= R/c$, decreases
again and finally shrinks to zero for $t=2R/c$. As in the case of the
thin sheet, the echo luminosity is approximately proportional to
$\tau_d\;R^{-1}$ and forward scattering has a similar effect.

\section{Numerical solution for a double exponential dust disk}
\label{sec:numeric}

The numerical integration of Eq.~\ref{eq:f} allows one to solve the
problem under the single scattering approximation for whichever dust
geometry and input SN light curve.

\begin{figure}
\centering
\resizebox{\hsize}{!}{\includegraphics{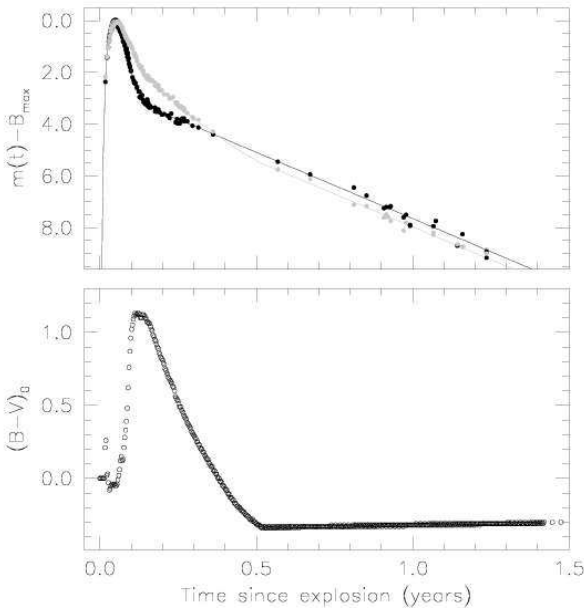}}
\caption{\label{fig:templates} Template light curves for $B$ (black)
and $V$ (gray) passbands.  The solid lines trace the smoothed version
used in the calculations while the filled dots indicate the observed
data. The lower panel shows the corresponding $(B-V)$ colour curve.  }
\end{figure}

For this purpose I have chosen to adopt a $V$ template light curve
obtained smoothing the combined data sets of SNe 1994D (\citealt{patat}, 
\citealt{richmond}, \citealt{capp97}) and 1992A (\citealt{suntzeff}, 
\citealt{capp97}), two very similar and well sampled
objects (see for example \citealt{patat}). The result is
shown in Fig.~\ref{fig:templates}, where I have plotted the
corresponding $(B-V)$ colour curve as well. The latter will serve to compare
the intrinsic SN colour with that of the LE (see Sec.~\ref{sec:colour}).

\begin{figure}
\centering
\resizebox{\hsize}{!}{\includegraphics{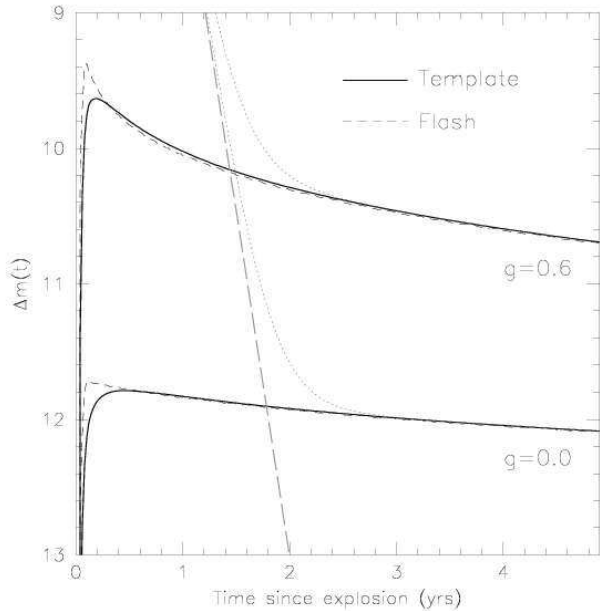}}
\caption{\label{fig:disk} Single scattering numerical solutions for a
face-on exponential disk for $g$=0 (lower curves) and $g$=0.6 (upper
curves).  The dotted curves trace the global light curves (SN+echo)
while the long dashed line indicates the radioactive tail of the
template light curve.  In all examples $C_{ext}$=5$\times$10$^{-22}$
cm$^2$, $\omega$=0.6, $R_{SN}$=2$\times$10$^4$ lyr and $Z_{SN}$=300
lyr were used (see text for more details).}
\end{figure}

In Fig.~\ref{fig:disk} I have traced the results obtained for a exponential
face-on dust disk, whose density profile was modeled using the typical
expression (see for example \citealt{bianchi}):

\begin{displaymath}
n(R,Z)=n_0 \; \exp \left (-\frac{R}{R_d}\right ) \; 
\exp \left( -\frac{|Z|}{Z_d} \right )
\end{displaymath}

where $R,Z$ are the cylindric galactic coordinates and $R_d,Z_d$ are
the characteristic radial and vertical scales. Following Bianchi et
al. (1996), I have adopted $R_d$=4.0 kpc and $Z_d$=0.14 kpc, which can
be considered as typical values for a spiral galaxy. The density $n_0$
is constrained by the central optical depth of the disk seen face-on,
which is given by $\tau(0)=2n_0 \; C_{ext} \; Z_d$.  Imposing a
typical value $\tau(0)$=1 (see \citealt{xilouris}) and for $C_{ext}$=
5$\times$10$^{22}$ cm$^2$, $n_0$ turns out to be 2.3 cm$^{-3}$. The SN
was placed at $R_{SN}$=2$\times$10$^4$ lyrs ($\sim$6 kpc) and
$Z_{SN}$=300 lyr above the galactic plane, which imply an optical
depth $\tau_d\approx$0.06.

For both isotropic and forward scattering I have run the calculations
using a flash (dashed line) and the template $V$ light curve
(thick solid line) and this shows that the results are pretty similar,
except for the early phases, where the flash tends to produce a
sharper and brighter peak. This is anyway irrelevant, since at such
epochs the SN is typically more than 10 magnitudes brighter that the
LE and therefore, at least for Ia's, Eq.~\ref{eq:flash} gives a fair
description of the real light curve.

The effect of forward scattering is clearly illustrated in
Fig.~\ref{fig:disk}: with $g$=0.6 the light curve starts to deviate
from the radioactive tail when the SN is about 1.5 mag brighter
than with $g$=0. Moreover, forward scattering has an effect on the
resulting slope for $t>$3 yrs, which is 0.1 and 0.05 mag yr$^{-1}$ in
the two cases respectively.

In conclusion, under rather normal conditions, a Type Ia SN exploding
in the disk of a spiral should always produce a LE which is of the
order of 10 mag fainter than the SN at maximum, without the SN being
heavily reddened.  Of course, placing the SN in the inner parts of the
disk or on the far side of the galaxy would increase the echo
luminosity and the extinction suffered by the SN itself. For example,
leaving all the other parameters unchanged and placing the SN at
$Z_{SN}$=0 would enhance the LE by 0.7 mag, while the optical depth would
grow to $\tau_d\sim$0.1, which is anyway still a rather low value.
Since there are good reasons to believe that $1\leq\tau(0)\leq 5$
(\citealt{bianchi04}), a Type Ia within 1-2 dust scale heights should
always produce an observable LE, unless it is located very far
from the galaxy center.

Using numerical integration, I have explored a few other simple cases,
like front sheets with different inclinations, thin spherical shells,
$r^{-2}$ winds and spherical clouds placed at different distances from
the SN and with varying offsets with respect to the line of sight. In
the latter case, when there is no scattering material along the line
of sight, the LE starts to take place at $t>$0 and, in this respect,
the parallel with acoustic physics becomes more pertinent, in the
sense that the reflected signal starts to be detached from the direct
impulse. Actually, in all cases where there is material on the line
of sight, one should rather talk about light reverbs (LRs), which
corresponds more appropriately to the analogous effect for acoustic
waves.

In general, it turns out that all dust geometries produce similar LE
luminosities and light curves. For typical values of $n_0$, $C_{ext}$,
$\omega$ and $g$, the LE at three years after the explosion has an
integrated brightness between 9 and 11 mag below SN
maximum. Therefore, unless one is able to resolve the LE image, it is
very difficult to disentangle between different dust geometries on the
basis of the light curve shape alone.

\section{Single scattering plus attenuation}
\label{sec:ssa}

\begin{figure}
\centering
\resizebox{\hsize}{!}{\includegraphics{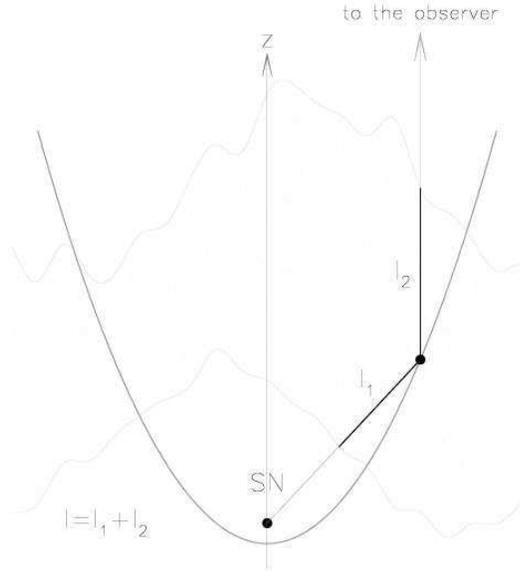}}
\caption{\label{fig:attgeom} Meaning of the optical path $l$ in the single 
scattering plus attenuation approximation.}
\end{figure}

The single scattering approximation can be refined in order to take
into account the presence of absorption. Some times this is referred
to as {\it single scattering plus attenuation} (hereafter SSA. See for
example \citealt{wood}). This approximation is obtained assuming that
the fraction of a photon packet which reaches the observer has
undergone one single scattering event on the iso-delay surface, while
the remaining fraction was scattered by a number of interactions with
the dust grains along the light path $l$ within the cloud system. The
path length $l$ is defined as the cumulated distance traveled from the
central source through the single scattering point and to the outer
boundary of the dust cloud in the direction of the observer across
regions with $n\neq0$ (see Fig.~\ref{fig:attgeom}). In the most
general case, $l$ depends on the $\varphi,z,t$ coordinates of the
scattering point (see Fig.~\ref{fig:geom} for the meaning of
$\varphi$). The inclusion of attenuation is achieved introducing a
$\exp{[-\tau_a(\varphi,z,t)]}$ factor in the definition of $f(t)$:

\begin{equation}
\label{eq:fssa}
f(t) = C_{ext} \; \omega\; \;c
\int_{-ct/2} ^{+\infty} \frac{\Phi(\theta)}{r}
\int_0^{2\pi} n(\varphi,z,t)\; e^{-\tau_a} \;d\varphi \; dz
\end{equation}

where the attenuation optical depth $\tau_a$ is defined as

\begin{equation}
\label{eq:taua}
\tau_a(\varphi,z,t)=C_{ext}\;\int_{l} n(\bmath{l})\; |d\bmath{l}|
\end{equation}

For one of the mean value theorems (see for example \citealt{rade}),
one can also write:

\begin{equation}
\label{eq:fssa2}
f(t) = \frac{C_{ext} \; \omega \; c}{2} \; e^{-\tau_{eff}}
\; G(t)
\end{equation}

where:

\begin{equation}
\label{eq:taueff}
e^{-\tau_{eff}}=\frac{\int_{-ct/2} ^{+\infty} \int_0^{2\pi} 
\frac{\Phi(\theta)}{r} \; n(\varphi,z,t)\; e^{-\tau_a(\varphi,z,t)} 
\;d\varphi \; dz}
{\int_{-ct/2} ^{+\infty} \int_0^{2\pi} 
\frac{\Phi(\theta)}{r}\; n(\varphi,z,t)\;d\varphi \; dz}
\end{equation}

and $G(t)$ is a time dependent function which includes the properties
of dust and dust distribution.  From Eq.~\ref{eq:fssa2} it is clear
that, with the introduction of attenuation, the overall LE luminosity
increases for growing density values until the optical depth reaches a
critical limit. After that, self-absorption prevails and any further
density enhancement causes the luminosity to decrease.

\begin{table}
\caption[]{Relevant parameters for the test-case geometries.}
         \label{tab:geom}
\begin{tabular}{lccc}
\hline \hline
Geometry          & $R$ $(R_1)$  & $\Delta R$ $(D,R_2)$ & $n_0(H)$ \\
                  & lyrs         & lyrs               & cm$^{-3}$\\
\hline
Sph. Shell     & 20          & 2               & 25, 125, 1000, 2500 \\
Sph. Cloud     & 50          & 100             & 0.5, 2, 20, 50 \\
Perp. Sheet    & 200         & 50              & 1, 5, 40, 100 \\
$1/r^2$ Wind   & 0.5         & 7              & 110, 530, 4250, 10700 \\
\hline
\end{tabular}
\end{table}

\begin{figure}
\centering
\resizebox{\hsize}{!}{\includegraphics{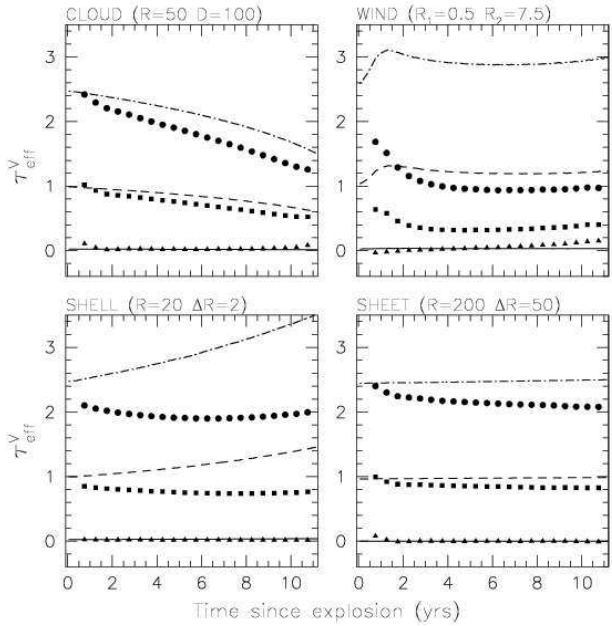}}
\caption{\label{fig:taue} Example $\tau_{eff}$ calculations for 
$\tau_d$=0.03 (solid line), 1.0 (dashed line) and 2.5 (dashed-dotted
line) in SSA approximation and $V$ passband. Geometrical parameters
are indicated in each panel. For comparison, filled symbols trace the
corresponding MC solutions (see Sec.~\ref{sec:mcres}).}
\end{figure}

In general, $\tau_{eff}$ is time and geometry dependent and it can be
regarded as a weighted optical depth for the LE at any given time. For
example, it is easy to demonstrate that for a SN immersed in a
perpendicular dust slab with uniform density, $\tau_a$ does not depend
on $\varphi$ and $z$ coordinates, since $l=D+ct$. Therefore the
solution of Eq.~\ref{eq:taueff} becomes straightforward and it gives
$\tau_{eff}=\tau_d + C_{ext} \;n_0 \;ct$, which implies that the
deviation from the single scattering approximation grows with time. A
few more realistic examples of $\tau_{eff}$ calculations for the $V$
band are presented in Fig.~\ref{fig:taue}, where I have considered
four different geometries: a uniform spherical shell extended between
$R$ and $R+\Delta R$, a distant perpendicular sheet placed at distance
$D$ and with a thickness $\Delta R$, a distant spherical cloud with a
radius $R$ and located at a distance $D$ in front of the SN and a
spherically symmetric wind with $n(r)=n_0\;(R_1/r)^2$, which extends
between $R_1$ and $R_2>R_1$. The first case can be considered as an
approximation to the so-called {\it contact discontinuity} (see
\citealt{sugerman03}). This can be produced by the mass loss undergone
by the Ia progenitor binary companion, through the interaction between
the slow dense red supergiant wind and the fast blue supergiant wind
(\citealt{emmering}). For the thin shell and perpendicular sheet I
have adopted values which are very similar to those of Sugerman (2003;
see his Table~4, second {\it B} case and first {\it I} case for SNe
respectively), for the spherical cloud I have used a typical size of 
molecular clouds systems (see for example \citealt{mathis00}), while
for the stellar wind I have allowed for an inner dust-free cavity
with radius $R_1$=0.5 lyrs ($\sim$5$\times$10$^{17}$ cm, see for example
\citealt{dwek}) and an outer radius $R_2/R_1$=15. As for the dust parameters,
I have adopted the values reported by \cite{weingartner} and \cite{draine}
for the $V$ passband (see also Table~\ref{tab:templates}).

For all geometries, the number densities were set in order to produce
the same optical depths $\tau_d$, i.e. the same SN extinction. The
relevant parameters are summarized in Table~\ref{tab:geom}. As one can
notice, for the stellar wind case the higher optical depths imply dust
densities which are rather unrealistic for this geometry and were
included for completeness, also following \cite{chev}.

As far as the relation between $\tau_{eff}$ and $\tau_d$ is concerned,
it must be noticed that this depends very much on the dust geometry.
For example, for a system with no dust along the line of sight,
$\tau_d$=0, while $\tau_{eff}$ can assume any value. This clearly
shows that, in general, the optical depth of a LE cannot be expressed
in terms of the optical depth of the SN as seen from the
observer. This is approximately true when the density is low or for
some particular geometries only. For example, for the slab case seen
before, $\tau_{eff}\simeq\tau_d$ for $ct\ll D$.  As we will see,
$\tau_{eff}$ actually depends not only on time and geometry, but also
on the dust properties.

\section{Monte Carlo simulations for multiple scattering} 
\label{sec:mc}

The idea behind Monte Carlo simulations is to follow each single
photon packet along its random walk until it finally escapes the dust
cloud. This procedure implicitly includes the treatment of multiple
scattering and it is therefore suitable for all cases where the dust
optical depth is large.

Following \citet{chev}, here I will adopt the procedure outlined by
\citet{witt}, recalling the fundamentals of the method and describing
in detail the aspects which are more specific to our case.

\subsection{Isotropic and beamed emission}
\label{sec:emission}

The whole process starts with the photon emission. Due to the
dimensions of the dust clouds I am going to consider, I can assume
that the SN is a point source lying at the origin of a coordinate
system $x,y,z$. If $\Re_i$ is a random number uniformly generated in
the range $0\leq \Re_i \leq 1$, the direction cosines are given by:

\begin{displaymath}
\begin{array}{lll}
w_0 & = & 2\Re_i-1 \\ u_0 & = & \sqrt{1-w_0^2}
\;\cos[\pi(2\Re_{i+1}-1)]\\ v_0 & = & \sqrt{1-w_0^2}
\;\sin[\pi(2\Re_{i+1}-1)]
\end{array}
\end{displaymath}

and obey to the usual condition $u_0^2+v_0^2+w_0^2$=1 (see
\citealt{witt}, Eqs.~5, 6 and 7).  When the scattering material, as
seen from the central source, shows significant voids (like in the
case of an isolated and distant cloud), the generation of random
photons on the whole unit sphere would be not only a waste of
computing time, but would also be inconsistent with the forced
scattering technique, which I will introduce in the next
sub-section. This problem can be solved emitting photons only within
the solid angle $\Omega_e$ subtended by the cloud at the SN position.
The fact that photons are artificially generated within a cone and not
on the whole unit sphere is taken into account during the
normalization of the light curve (see Sec.~\ref{sec:lc},
Eq.~\ref{eq:lcnorm}).

\subsection{Forced first scattering}
\label{sec:forced}

To avoid wasting time with photons directly escaping the cloud, I
have followed the common practice using the so called {\it forced
first scattering} (\citealt{witt}). The basic idea behind this
technique is to scatter each emitted photon, while the direct escape
of non interacting photons is then taken into account using a
weighting factor $W_0=1-\exp^{-\tau_1}$.

Once the random direction is generated, one needs to compute the total
optical depth along the light path $\bmath{r}$:

\begin{displaymath}
\tau_1=\int_0^{r_{max}} n(\bmath{r}) \; C_{ext} \; dr
\end{displaymath}

where $r_{max}$ is the maximum dimension of the cloud or system of
clouds along the emission direction.  The next step is the random
generation of $\tau\leq \tau_1$ at which the first scattering will
occur. This is achieved using $e^{-\tau}$ as the probability density
function, applying the inverse method and solving for $\tau$. This gives
the following expression:

\begin{displaymath}
\tau=-\ln\left [ 1-\Re(1-e^{-\tau_1})\right ]
\end{displaymath}

which corresponds to Eq.~14 of \citet{witt}, except from the
fact that the latter has the incorrect sign.

Once this is done, one needs to determine the spatial position
$\bmath{r}$ where the scattering occurs. For simple dust geometries
this can be obtained inverting the analytical relation which links
$\tau$ to the particle density function, while more general cases can
be solved numerically (see for example \citealt{fischer}), at the
expense of computing time.

Irrespective of the adopted method, one has then to compute the photon
direction after the scattering event, in order to be able to decide
about its immediate destiny. For this purpose, as in the case of the
analytical solutions, I have adopted the H-G scattering phase
function. Again, using the inverse method, and solving for $\mu_1$ one
finally obtains:

\begin{equation}
\label{eq:mu1}
\mu_1=\frac{1}{2g}
\left [
(1+g^2)-\frac{(1-g^2)^2}{(1+2g\Re-g)^2}
\right ]
\end{equation}

which corresponds to Eq.~19 of \citet{witt}, except from the fact that
the exponent 2 on the $(1-g^2)$ term was lost. For $g=0$,
Eq.~\ref{eq:mu1} has to be replaced by $\mu_1=2\Re-1$.

Once $\mu_1$ is generated according to Eq.~\ref{eq:mu1}, the new
direction cosines $(u_1, v_1, w_1)$ need to be computed. This is done
by the following coordinate transformation (see for example
\citealt{bianchi}, Appendix A4):

\begin{displaymath}
\begin{array}{l}
u_1 = u_0 \cos\theta -\frac{v_0}{\sqrt{1-w_0^2}}\sin\theta\sin\phi\;
+\frac{u_0w_0}{\sqrt{1-w_0^2}}\sin\theta\cos\phi\\ 
v_1 = v_0 \cos\theta -\frac{u_0}{\sqrt{1-w_0^2}}\sin\theta\sin\phi\;
+\frac{v_0w_0}{\sqrt{1-w_0^2}}\sin\theta\cos\phi\\ 
w_1 = w_0 \cos\theta- \sqrt{1-w_0^2}\sin\theta\cos\phi\\
\end{array}
\end{displaymath}

where $\theta$ is the scattering angle defined by $\mu_1=\cos\theta$,
and $\phi$ is the scattering azimuth, from which I assume the
scattering efficiency to be independent\footnote{Note that the
equivalent Eqs.~22 given by \cite{witt} are affected by several
typos.}. This is generated simply as $\phi=\pi\;(2\Re-1)$, which gives
a uniform distribution in the range $-\pi\leq \phi \leq \pi$.

\subsection{Successive scatterings and final photon escape}
\label{sec:succscatt}

The optical path to the next possible scattering can be computed with
considerations which are similar to those I made for the first
scattering, except from the fact that now photons are not forced to
interact with the dust. Under this condition, the optical depth at
which the next scattering event will occur is simply given by
$\tau=-\ln \Re$.  Whether the photon will escape or undergo further
scatterings can be decided comparing $\tau$ with $\tau_2$, which is now
computed along the scattered direction from the scattering point to
$r_{max}$. If $\tau>\tau_2$ the photon will escape and, to compensate
for the first forced scattering, a weight $W=\omega \; W_0$ will be
given to it. The dust albedo $\omega<1$ accounts for the fact that
during the scattering some photons can actually be absorbed.

In the case of $\tau\leq \tau_2$ the photon packet will be scattered
again and hence the whole process has to be repeated, starting with
the calculation of the scattering point, scattering direction and so
on, until it finally escapes. In the general case of $N\geq 1$
scattering events, the final weight will be 

\begin{equation}
\label{eq:weight}
W=\omega^{N}\;W_0
\end{equation}

Once the emerging fraction of a given photon packet escapes, there are
a series of parameters that are relevant for its temporal and spatial
classification: a) the position $P_N(x_N, y_N, z_N)$ of last
scattering; b) the direction cosines after last scattering $(u_N, v_N,
w_N)$; c) the time delay accumulated by the scattered photons with
respect to those which were emitted at the same time but traveled
directly to the observer. All these numbers are needed to
reconstruct integrated luminosity and surface brightness image (or
profile) at a given time and from a given viewing direction.

While parameters a) and b) are immediately available, the calculation
of time delay requires a little bit of more discussion.

\subsection{Light curve reconstruction}
\label{sec:lc}

Escaping photons can be seen only by an observer who lies on the line
of sight coinciding with the last scattering direction $\bmath{p}(u_N,
v_N, w_N)$.  If $\bmath{d}$ is the vector which goes from the coordinate
system origin to the observer, $\bmath{r}(x_N, y_N, z_N)$ defines the
position of last scattering $P_N$ and $\bmath{l}=l\bmath{p}$ is the vector
which goes from $P_N$ to the observer, one can clearly write
$\bmath{d}=\bmath{r}+\bmath{l}$.  Then if I set $k=\bmath{r}\circ\bmath{p}
\equiv(x_Nu_N+y_Nv_N+z_Nw_N)$, I obtain $d^2=r^2+2kl+l^2$, which can
be solved for $l$. One of the two solutions of this equation is not
physical and therefore the light path from the last scattering
to the observer is $l=-k+\sqrt{k^2+d^2-r^2}$, where
$r^2=x_N^2+y_N^2+z_N^2$. For $d\gg r$ this gives $l\approx
d-k$. Taking into account the random walk which was traveled before
the last scattering, one finds that the total optical path is
$L=l+\sum_{i=1}^N r_i$. Having this in mind, the arrival time $t_a$
can be easily computed as:

\begin{equation}
\label{eq:ta}
t_a = t_e + \sum_{i=1}^N \frac{r_i}{c} \; + \;\frac{l-d}{c}
\simeq \sum_{i=1}^N \frac{r_i}{c}\;- \;\frac{k}{c}
\end{equation}

where $t_e$ is the emission time. This has to be generated according
to the input SN light curve $L(t)$, which in this case can be regarded
as a probability density function. Applying the inverse method I can
write:

\begin{displaymath}
\Re=\frac{1}{L_0\;\Delta t_{SN}} \int_0^{t_e} L(t)\; dt
\end{displaymath}

Since, in principle, one does not have an analytical description of $L(t)$,
the integral has to be evaluated numerically for a series of $t_e$
values and the result can be easily interpolated and inverted, so that
entering the uniform random variable $\Re$ one gets $t_e$ with the
correct distribution. Of course, in the case of flash approximation
(see Eq.~\ref{eq:flash}), the emission time is simply generated as
$t_e=\Delta t_{SN}\; \Re$.

At this point, the light curve can be progressively computed just
accumulating the weighted photon packets in time delay bins $\Delta T$
along a given direction and within a given solid angle $\Delta
\Omega$. Now, if $\bmath{p}(u_N,v_N,w_N)$ is the photon escaping
direction and $\bmath{s}(s_x,s_y,s_z)$ is the versor which identifies
the line of sight, the angle $\psi$ formed by the two vectors is
defined by $\cos\psi = \bmath{p} \circ \bmath{s}$. Photons are
counted if $\cos\psi\geq$cos$\psi_l$, where the limiting angle can be
defined as $\cos\psi_l=1-q$ and $\Delta \Omega=2\pi q$, with $0\leq q
\leq 1$.

The bin size $\Delta T$, which in the end will give the time
resolution, depends on the total number of generated packets and the
signal-to-noise ratio one wants to achieve. The same applies to
$\Delta \Omega$, which will govern the angular resolution.

Finally, in order to follow the same procedure I adopted for the
analytical solutions, the resulting light curve needs to be normalized
to the SN maximum light luminosity. This is achieved dividing it by
the number of photons $F_0$ the observer would receive by the SN at
maximum, within the solid angle $\Delta \Omega$ and in the time
interval $\Delta T$, which is clearly given by

\begin{equation}
\label{eq:lcnorm}
F_0=N_{phot}\;\frac{\Delta T}{\Delta t_{SN}}\; \frac{\Delta
\Omega}{\Omega_e}
\end{equation}

For spherically symmetric models, escaping photons can be both
generated and collected all over the unit sphere ($\Delta \Omega =
\Omega_e =4\pi)$, and hence the normalization is obtained dividing the
photon counts by $N_{phot}\;\Delta T/\Delta t_{SN}$.

\subsection{Echo surface brightness reconstruction}
\label{sec:surface}

Escaping direction cosines and last scattering position can be used to
derive the spatial energy distribution projected onto the plane
perpendicular to the line of sight, which coincides with the plane of
the sky. This can be achieved simply transforming the coordinates of
the last scattering point $P_N(x_N,y_N,z_N)$ in the new system
$O(X,Y,Z)$, which is obtained rotating by an angle $\beta$ around the
$z$-axis and by an angle $\gamma$ around the transformed $y$-axis. The
rotation angles $\beta$ and $\gamma$ are defined by the escaping
direction cosines as:

\begin{equation}
\label{eq:betagamma}
\begin{array}{ll}
\sin\beta = v_0/\sqrt{1-w_0^2}; &
\cos\beta = u_0/\sqrt{1-w_0^2} \\
	& \\
\sin\gamma  = w_0; &
\cos\gamma  = \sqrt{1-w_0^2}
\end{array}
\end{equation}

In this reference system, the plane of the sky coincides with the
$Y,Z$ plane, while the $X$-axis coincides with the line of sight. The
transformation equations are as follows:

\begin{displaymath}
\begin{array}{l}
X = x_N \cos\alpha \cos\beta + y_N\sin\alpha \cos\beta +
z_N\sin\beta\\ Y = -x_N \sin\beta + y_N \cos\beta\\ Z = -x_N \cos\beta
\sin\gamma - y_N \sin\beta \sin\gamma + z_N \cos\gamma
\end{array}
\end{displaymath}

Of course, contrarily to what happens for the numerical integration of
Eq.~\ref{eq:surfbr}, MC simulations do not allow direct calculation of
the echo surface brightness at a given time $t$, since the arrival
times $t_a$ are scattered across the whole range allowed by the dust
geometry.  For this reason, one needs to perform a selection on the
outcoming photons and collect them only if their arrival times
satisfy the condition $t-\Delta T/2 \leq t_a \leq t+\Delta T/2$. Due
to the large difference between the overall echo duration and the time
resolution $\Delta T$, this means that only a small fraction of the
total number of generated photons will fall in the required time
range.

In order to achieve a reasonable signal to noise ratio in the final
results, this implies that a very large number (10$^8-$10$^9$) of
photons must be generated.

\section{Results of MC simulations}
\label{sec:mcres}

I have implemented the concepts discussed in the previous sections in
a MC code, which I have used to study the effects of multiple
scattering on the outcoming LEs for the four dust geometries described
in Sec.~\ref{sec:ssa}.  In all calculations I have adopted
$C_{ext}$=5.21$\times$10$^{-22}$ cm$^2$, $g$=0.54 and $\omega$=0.66,
which are typical of a canonical $R_V$=3.1 Milky Way mixture in the
$V$ passband (see \citealt{draine} and Sec.~\ref{sec:spectrum}
here). For non spherically symmetric geometries, I have used $\Delta
\Omega$= 2$\times$10$^{-3} \pi$ ($q$=10$^{-3}$), which corresponds to
a collecting beam with an angular semi-amplitude of $\sim$2.6 degrees.

\begin{figure*}
\centering
\resizebox{\hsize}{!}{\includegraphics{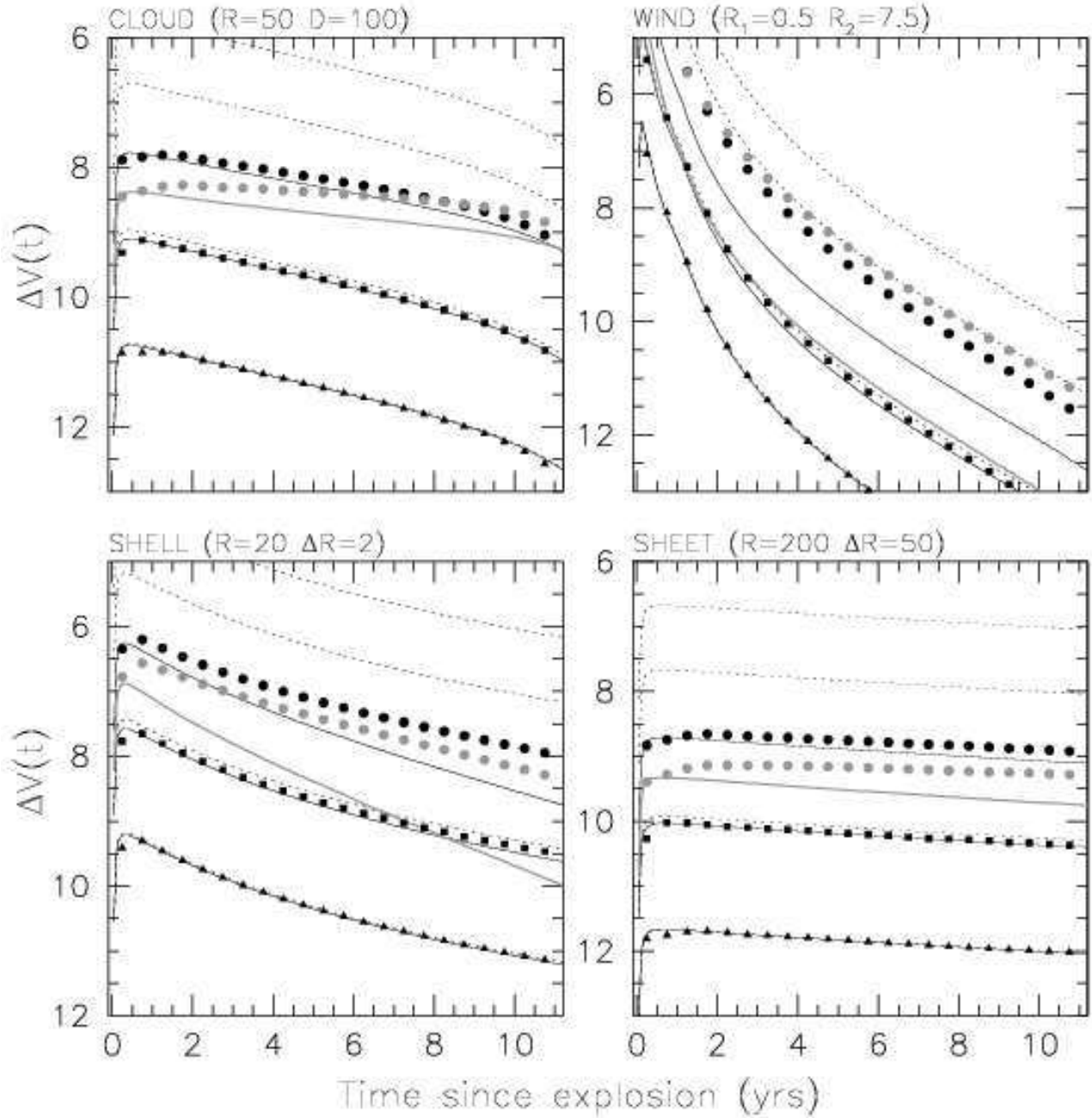}}
\caption{\label{fig:mcres}MC solutions for four different dust geometries
for $\tau_d$=0.03 (triangles), 0.1 (squares), 1.0 (circles) and 2.5
(gray circles).  For comparison, single scattering and SSA solutions
are also indicated (dotted and solid lines respectively). For clarity,
the SSA solution for $\tau_d$=2.5 has been traced with a thick gray line.}
\end{figure*}

The basic effects of multiple scattering are illustrated in
Fig.~\ref{fig:mcres} for different values of $\tau_d$.  While the MC
simulated luminosity is only a few percents fainter than the single
scattering solution in the low optical depth case ($\tau_d$=0.03),
this difference becomes more and more marked as the dust density
grows. For $\tau_d$=0.1 the deviation is $\sim$0.1 mag and it reaches
about 1 mag for $\tau_d$=1.0. This is in very good agreement with the
results found by \citet{chev} in the case of type II SNe surrounded by
dust distributed with a $r^{-2}$ density law.

This is clearly due to the growth in the number of multiple
scatterings. In fact, while the number of single scatterings tends to
increase the echo luminosity, further scatterings contribute to the
auto-absorption and therefore work against any brightness increase.
As a consequence of these two competing mechanisms, the echo
luminosity keeps growing for increasing values of $n$ until the
fraction of photons which undergo pure single scatterings reaches
0.5. The scattered flux increases more and more slowly as the dust
optical depth approaches some critical value ($\tau_d\simeq$1), after
which the fraction of photons with $N_{scat}\geq2$ becomes larger than
0.5 and the overall luminosity decreases for any further increase of
$\tau_d$. The effect of a further dust density rise is illustrated by
the gray circles, which were obtained for $\tau_d$=2.5. The exact
details depend on the specific dust distribution but, for example, in
the case of the thin shell, under these circumstances $N_{scat}$=1 in
$\sim$15\% of the cases only and this turns into an overall echo
luminosity which is more than 1.5 mag fainter than the corresponding
single scattering solution. Very similar results are obtained for the
distant sheet and the $r^{-2}$ cases, while for the spherical cloud
the behaviour is different, due to the fact that the intersection
between the LE paraboloid and the cloud tends to rapidly decrease.

While this is not foreseen in the plain single scattering
approximation, this is qualitatively predicted by the SSA
approximation (solid curves), which gives reasonably good results for
$\tau_{eff}<1$. In general, the simulations show that SSA successfully
reproduces the early phases under a wide range of geometries, while it
progressively deviates from the MC solutions as time goes by, and the
deviation is larger for higher values of $\tau_{eff}$.  The light
curves produced by SSA are systematically fainter than multiple
scattering solutions. This is a consequence of the assumption on which
SSA is based, i.e. that attenuation acts as a pure absorption, which
is approximately true at low optical depths only. In fact, when the
density grows, photons which are deviated by the single scattering
trajectory still have a probability of reaching the observer through
further scatterings and can, therefore, contribute to the LE at later
times.  This effect, already noticed by \citet{chev}, translates into
a light curve flattening and it can be seen as the consequence of an
increased average time delay accumulated by photons. The net effect
is to reduce the effective LE optical depth with respect to
$\tau_{eff}$ implied by the SSA approximation (see
Fig.~\ref{fig:taue}, filled symbols). The simulations show that this
is more effective when the dust is confined closer to the SN and, for
a given $\tau_{eff}$, when the dust density is higher.

More subtle is the outcome of multiple scattering on the radial
surface brightness profile. For a uniform dust distribution, the
simulations show that the resolved LE would always appear as a disk
with a small surface brightness increase going outward and the
gradient tends to become more pronounced for larger values of the dust
optical depth. However, the effect is practically negligible.

Finally, the dust optical depth causes a progressive breaking of the
relation between $r$ and $\rho$ expressed by Eq.~\ref{eq:eqdist}
which, in principle, allows the direct calculation of the dust
distance from the SN from the resolved LE image. In fact, in the high
density regime the contribution by multiple scatterings tends to fill
all the region internal to the iso-delay surface, mimicking the
presence of dust close to the SN itself. I must notice, however, that
the effect is not really pronounced and in most cases it would not be
observable, the multiple scattering contribution being lost in the
background noise.

\section{Light echo spectrum}
\label{sec:spectrum} 

So far I have restricted the discussion to one single wavelength. But,
in principle, a LE can be studied at any wavelength,
provided that the dependencies of the relevant parameters, namely
$C_{ext}$, $\omega$ and $g$, are known. If this is the case, the light
echo spectrum can be computed using Eq.~\ref{eq:F}, where $L(t)$ is
replaced by the spectrum $S(\lambda,t)$.

\begin{figure}
\centering
\resizebox{\hsize}{!}{\includegraphics{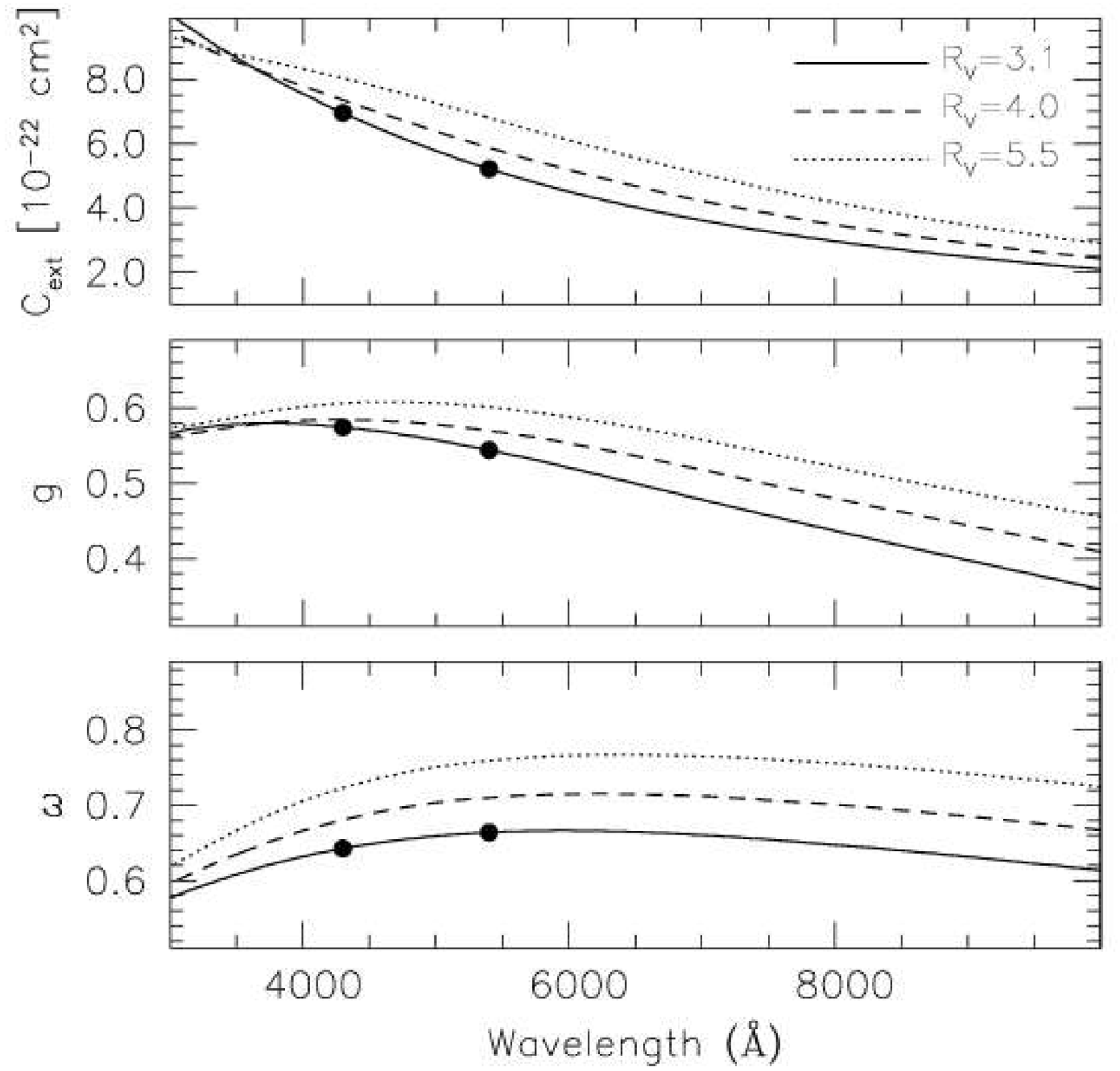}}
\caption{\label{fig:draine} Behaviour of albedo (lower panel) forward
scattering degree (middle panel) and extinction cross section (upper
panel) for a Milky Way size dust distribution (\citealt{weingartner},
\citealt{draine}) and different values of $R_V$. Filled dots indicate
the values at the effective wavelengths of $B$ and $V$ passbands.}
\end{figure}

For this purpose I have adopted the results published by
\cite{weingartner} and \cite{draine} for a Milky Way 
mixture of carbonaceous and silicate grains with a canonical
$R_V$=3.1. The behaviour of the relevant parameters in the optical
spectral range is shown in Fig.~\ref{fig:draine} where, for
comparison, I have also plotted the functions computed for higher
values of $R_V$. As one can see, the dust albedo changes very little
across the optical range (0.60$\leq\omega\leq$0.67), while the forward
scattering degree shows a more significant variation. However, the
strongest wavelength dependency is shown by the extinction
cross-section whose behaviour, in the wavelength range 4000-9000 \AA,
is well approximated by a $\lambda^{-\alpha}$ law, with $\alpha$=1.35.

These functions were implemented in the codes by means of polynomial
interpolations to the tabulated values\footnote{Tabulated functions
can be found at the following URL: {\tt
http://www.astro.princeton.edu/$\sim$draine/dust/dustmix.html}}.

\subsection{Single scattering}
\label{sec:sinspec}

In single scattering approximation, as in the case of the light curve,
the echo spectrum \ shape is independent of $n_0$, which governs the
global luminosity only, while dust geometry and dust properties
effects are included through the impulse response function $f(t)$.
Now, the SN luminosity decays very fast and thus the bulk of optical
radiation is emitted in a very short time range\footnote{For a normal
Ia, 90\% of the radiation between 3700~\AA\/ and 7200~\AA\/ is
released in about 0.3 yrs.}, which I will indicate as $\Delta t_b$.
As a consequence, what really counts for the shape of the output
spectrum observed at time $t$ is the behaviour of $f$ between
$t-\Delta t_b$ and $t$.  Therefore, the geometrical dependency is
expected to be generally very mild at $t\gg\Delta t_b$.  Exceptions to
this might be encountered, for example, during the onset of a real
echo, in which $f$ shows a pronounced transient phase.  For the same
reason, also variations on the forward scattering degree $g$ are
expected to produce significant changes in the overall flux, but not
in the spectral shape.

\begin{figure}
\centering
\resizebox{\hsize}{!}{\includegraphics{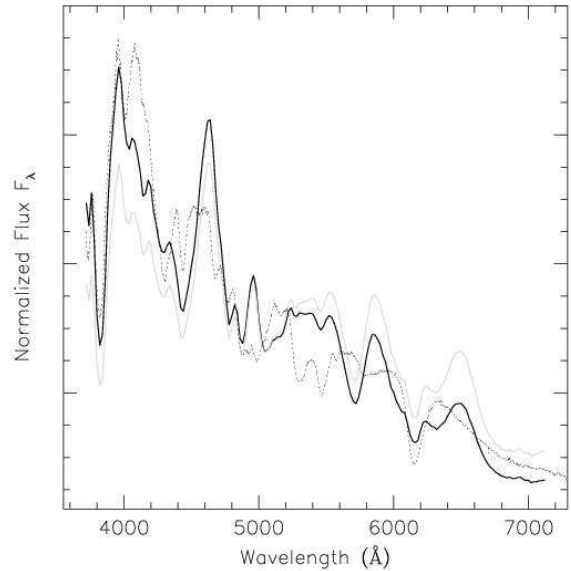}}
\caption{\label{fig:spectrum} Light echo spectrum computed at $t$=2 yrs in 
single scattering approximation for a thin shell with $R$=20 lyr,
$\Delta R$=2 lyr (black line) and $\tau_d\approx \tau_{eff}$=0.03
($n_0$=25 cm$^{-2}$). For comparison, the solid gray line and the
dotted line trace the time integrated spectrum and the maximum light
spectrum of SN~1998aq (Branch et al. 2003), respectively. For
presentation all spectra have been normalized to their integrated
luminosity in the wavelength range 3720-7160~\AA.}
\end{figure}

This is confirmed by the numerical solutions, for which I have used
the spectroscopic data set of SN~1998aq \citep{branch}. This includes
29 spectra obtained in the phase range $-$9$^d\leq t \leq +$241$^d$
and covers the wavelength range 3720-7160~\AA.  An example of LE
spectrum is shown in Fig.~\ref{fig:spectrum}: it is immediately clear
that it does not resemble the SN spectrum at maximum, even if it has
roughly the same colour (see also next section). For comparison, in
Fig.~\ref{fig:spectrum} I have also plotted the time integrated
spectrum of SN~1998aq (gray line). This is clearly redder than the
actual LE spectrum, mainly due to the wavelength dependency of
$C_{ext}$.

As I have already pointed out, in single scattering approximation an
increase of the dust density produces a proportional increase in the
overall echo flux, but no variation in its spectral appearance. In
order to investigate optical depth effects, one in fact needs to use
the multiple scattering approach. This is of course possible, even
though the number of photons required to reach a reasonable signal to
noise ratio in the output spectrum becomes very large. Practically
this allows echo spectral synthesis in spherical symmetry only and,
therefore, geometrical effects must be evaluated using broad band
colours instead (see next section).

\subsection{Multiple scattering}
\label{sec:mulspec}

The procedure to compute a multiple scattering spectrum is identical
to the one I have described for the light curve, with the exception
that now the emission time $t_e$ and the emitted wavelength
$\lambda_e$ have to be generated according to the input spectral
distribution $S(\lambda,t)$. This can be easily achieved using the
inverse theorem. Once the photon has completed its random walk through
the dust, it is finally classified according to the arrival time $t_a$
and counted in the proper wavelength bin, whose amplitude $\Delta
\lambda$ fixes the spectral resolution.

\begin{figure}
\centering
\resizebox{\hsize}{!}{\includegraphics{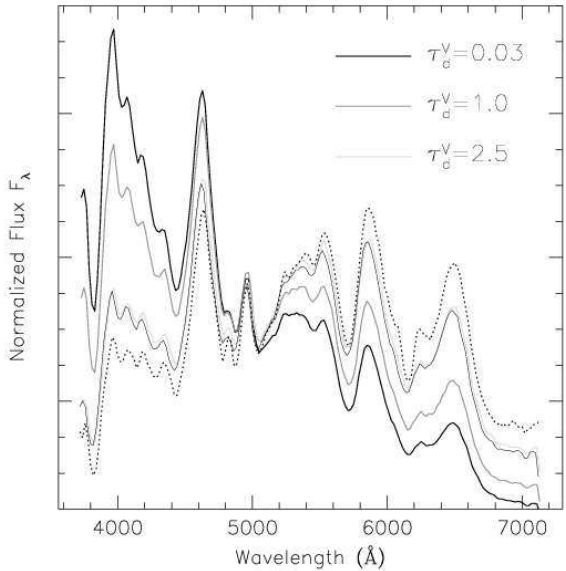}}
\caption{\label{fig:mcspec} Synthetic LE spectra for a thin spherical shell
($R$=20 lyr, $\Delta R$=2 lyr) at $t$=2 yrs for increasing values of
$\tau^V_d$ (dark to light thick lines). The thin line is the low
optical depth spectrum reddened with a suitable extinction (see text);
for clarity the spectrum was slightly shifted downward.  For
comparison, the dotted line traces a spectrum computed for
$\tau^V_d$=2.5 with the SSA approximation. For presentation all
spectra have been normalized to their integrated luminosity in the
wavelength range 3720-7160~\AA.}
\end{figure}

The results for a thin shell ($R$=20 lyr, $\Delta R$=2 lyr) and $t$=2
yrs are shown in Fig.~\ref{fig:mcspec}, for three different values of
$\tau_d$.  As expected, the introduction of multiple scattering
produces significant reddening in the spectrum. As a matter of fact,
the spectra calculated for higher densities can be obtained directly
from the one computed for the lowest $\tau_d$, applying the same
wavelength dependency of $C_{ext}$ used for the LE calculation and a
suitable optical depth $\tau_{eff}$. An example is presented in
Fig.~\ref{fig:mcspec} (solid thin line), which was obtained using
$\tau^V_{eff}$=1.95, i.e. the value derived from multiple scattering
light curve model (see Fig.~\ref{fig:taue}). For comparison, in 
Fig.~\ref{fig:mcspec} I have also plotted the spectrum obtained in SSA 
approximation for $\tau^V_d=2.5$ (dotted line). This appears to be
too red with respect to the corresponding MC solution and the reason is that
SSA tends to overestimate the LE optical depth by amounts which become
larger at shorter wavelengths (see also next section).

\section{Light echo colour}
\label{sec:colour} 

An alternative to spectral synthesis is the much less expensive
calculation of broad band light curves, which can be combined to
produce synthetic colours. In order to explore the potentialities of
this method, I have studied the case of the $(B-V)$ colour, since
for a Ia the LE is clearly brighter in these passbands than at redder
wavelengths (see Fig.~\ref{fig:spectrum}).

\begin{table}
\caption[]{Basic parameters for $B, V$ light curves and dust mixtures.
Values for $C_{ext}$, $\omega$ and $g$ are from \cite{weingartner} and
\cite{draine} for $R_V$=3.1 (first two rows) and $R_V$=5.5 (last two
rows). Extinction cross section $C_{ext}$ is given per Hydrogen atom.}
\label{tab:templates}
\begin{tabular}{cccccc}
\hline \hline
Passband & $\lambda$ & $C_{ext}$ & $\omega$ & $g$ & $\Delta t_{SN}$ \\ 
         & (\AA)     &(10$^{-22}$cm$^2$)  &          &     & (yrs) \\
\hline
         &      &                        &      &      &   \\ 
B        & 4300 & 6.95                   & 0.64 & 0.57 & 0.0654\\ 
V        & 5400 & 5.21                   & 0.66 & 0.54 & 0.0857\\
         &      &                        &      &      &       \\
B        &      & 8.03                   & 0.72 & 0.61 &       \\
V        &      & 6.80                   & 0.76 & 0.60 &       \\
\hline
\end{tabular}
\end{table}

\subsection{Single scattering plus attenuation}
\label{sec:attenuation} 

A first analytical estimate can be derived in SSA approximation,
assuming that the input light curve is a flash. According to what I
have shown in Secs.~\ref{sec:single} and \ref{sec:ssa}, under this
assumption the echo flux at a given wavelength is $F= L_0\; n_0\;
C_{ext}\;
\Delta t_{SN} \; e^{-\tau_{eff}} \;G(t)$, where $G(t)$ is a time and 
wavelength dependent function related to geometry and dust
properties. Applying this to the case of $B$ and $V$ passbands, 
and indicating with $(B-V)_{max}^0$ the unreddened colour of the SN at
maximum, one can estimate the LE colour as:

\begin{eqnarray}
\label{eq:splusabs}
(B-V) & \approx & (B-V)_{max}^0 - 2.5\;\log\;\frac{C_{ext}^B \;\omega^B \Delta
t_{SN}^B}{C_{ext}^V \; \omega^V \; \Delta t_{SN}^V }\nonumber\\
      &  +      & 1.086\;(\tau^B_{eff}-\tau^V_{eff})
\end{eqnarray}

where, due to the mild wavelength dependency of $g$ (see
Table~\ref{tab:templates}), I have assumed $G^B(t)\approx G^V(t)$.  For
low values of $\tau_{eff}$, after substituting the relevant parameters
in Eq.~\ref{eq:splusabs}, one gets
$(B-V)\approx(B-V)_{max}^0\simeq-$0.1. Therefore, when the single
scattering approximation is valid, the LE colour is roughly time and
geometry independent. More generally, when the LE optical depth is not
negligible, Eq.~\ref{eq:splusabs} implies that the resulting colour is
reddened by auto-absorption ($\tau^B_{eff}>\tau^V_{eff}$). Moreover,
it is directly related to the LE optical depth. In fact, using the
values reported in Table~\ref{tab:templates}, Eq.~\ref{eq:splusabs} can
be rewritten as:

\begin{equation}
\label{eq:bmvappr}
(B-V)\approx  (B-V)_{max}^0 + 0.36 \;\tau^V_{eff}
\end{equation}

Thus, the observed $(B-V)$ colour can be used to estimate the
effective optical depth of a LE, so that this is not a free
parameter. Actually, numerical simulations show that the ratio
$G^B(t)/G^V(t)$ can vary with time and geometries between 0.9 and 1.1,
so that the previous simplified equation is expected to produce
maximum errors of about $\pm$0.1 on the derived $(B-V)$ colour.

\subsection{Multiple scattering}

As we have seen in Sec.~\ref{sec:mcres}, when multiple scattering is
at work, the echo luminosity does not scale linearly with the density and
therefore the observed colour is expected to show a dependency on the
dust optical depth. In particular, for increasing values of particle
density, light curves appear to get fainter and fainter with respect
to the corresponding single scattering solutions (see for example
Fig.~\ref{fig:mcres}). Since this effect is supposed to get stronger
for higher values of $C_{ext}$ (i.e. at shorter wavelengths), one
expects the echo colour to become redder and redder as the density
grows, as qualitatively foreseen also in the SSA approximation.

\begin{figure}
\centering
\resizebox{\hsize}{!}{\includegraphics{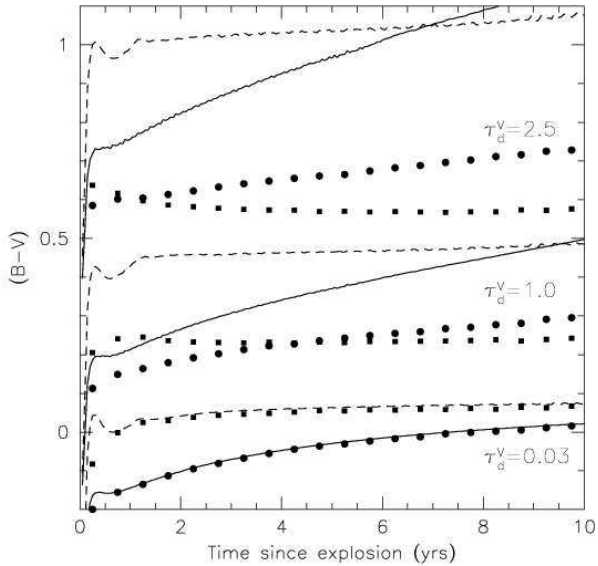}}
\caption{\label{fig:bmv} Multiple scattering $(B-V)$ colour curves for the 
thin shell (circles) and $r^{-2}$ wind (squares) for three different values
of $\tau_d$. The corresponding SSA solutions are traced by solid and
dashed curves respectively.}
\end{figure}

\begin{figure}
\centering
\resizebox{\hsize}{!}{\includegraphics{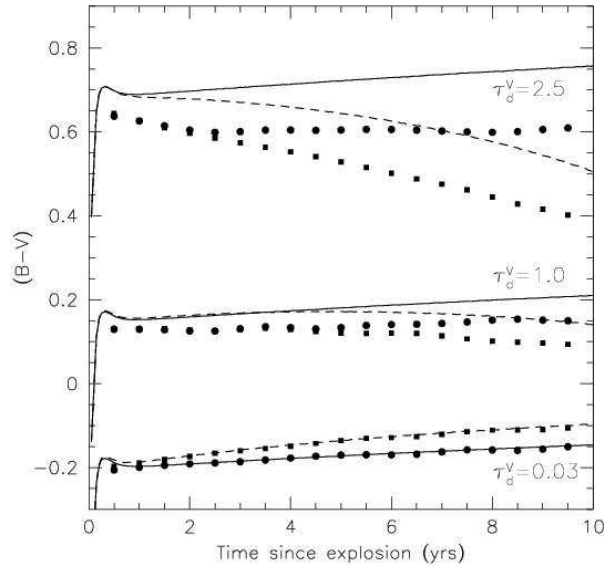}}
\caption{\label{fig:bmvother} Multiple scattering $(B-V)$ colour curves for 
the perpendicular sheet (circles) and spherical cloud (squares) for
three different values of $\tau_d$. The corresponding SSA solutions are 
traced by solid and dashed curves respectively.}
\end{figure}

To explore these ideas in a more quantitative way, I have run a series
of calculations using the parameters listed in
Table~\ref{tab:templates}, which were derived from the template light
curves presented in Fig.~\ref{fig:templates} and from Draine's models.
The results for the thin shell and the $r^{-2}$ wind are shown in
Fig.~\ref{fig:bmv}, for three different values of $\tau^V_d$, while
Fig.~\ref{fig:bmvother} presents the cases of the distant sheet and the
spherical cloud. For low optical depths the MC solutions are
practically identical to the SSA solutions (solid and dashed curves)
while, for larger densities, the calculated colour curves turn
globally redder and redder for both geometries and progressively
deviate from the SSA solutions. In general, these latter are
systematically redder than the multiple scattering solutions and they
give reasonable results only for $\tau_{eff}<$1.

As far as the time dependence is concerned, the degree of forward
scattering plays an important role. In fact, the random walk of a
photon through the medium is influenced by the scattering phase
function, in the sense that for high $g$ values, a photon tends to
follow its emission direction, since small scattering angles are
favored, and therefore its random walk will be shorter than for lower
$g$ values.  This means that the light curve is expected to declines
faster at bluer wavelengths, where $g$ is higher (see
Fig.~\ref{fig:draine}).  On the other hand, since $C_{ext}$ also
increases bluewards, this produces the opposite effect, implying a
light curve flattening, which is more efficient in the blue. As the
simulations show, for the Milky Way mixture we have adopted
(\citealt{weingartner}, \citealt{draine}) the two mechanisms tend
almost to compensate each other, giving a rather flat time evolution,
mostly driven by geometrical effects. As a matter of fact, the exact
behaviour turns out to be pretty sensible to the wavelength dependency
of $g$. Both SSA and MC calculation show that if one assumes $\partial
g /\partial \lambda$=0, the $(B-V)$ colour tends to become bluer with
time. Of course, the time evolution will also depend on the dust
distribution, mainly through the allowed incoming angle range
variation and the changes in $\tau_{eff}$. For example, when the dust
is far from the SN, as in the case of the perpendicular sheet
(Fig.~\ref{fig:bmvother}, circles), this does not vary significantly
with time, and therefore the colour curve is expected to be flatter
than for dust close to the central source. This is also true for the
distant spherical cloud (Fig.~\ref{fig:bmvother}, squares), but the
fact that $\tau_{eff}$ decreases as time goes by causes the LE to
become progressively bluer.

The main conclusion is that the LE colour is practically time and geometry
independent, especially after the initial onset phase ($t>$2 yrs), provided
that the dust density is uniform. Therefore, any significant change in the
observed colour has to be interpreted as a variation of $\tau_{eff}$ which,
in turn, would signal the presence of density fluctuations.

As in the case of SSA approximation, one can use the approximate
expression given by Eq.~\ref{eq:bmvappr} to estimate the effective
multiple scattering optical depth directly from the observed $(B-V)$
colour. It must be noticed that the assumption that
$\tau^B_{eff}/\tau^V_{eff}\equiv C^B_{ext}/C^V_{ext}$, which holds in
SSA approximation, is not strictly true when multiple scattering is at
work. In general, in fact, this depends on the particular geometry and
tends to be slightly larger than the cross sections
ratio. Nevertheless, the simulations show that Eq.~\ref{eq:bmvappr}
gives reasonable results for $\tau^V_{eff}\leq2$, where now
$e^{-\tau_{eff}}$ is defined as the ratio between the multiple
scattering solution and the corresponding single scattering solution.

\section{Effects of a dust mixture change}
\label{sec:mixture} 

In all calculations, I have so far adopted a canonical $R_V$=3.1 Milky
Way dust composition. To illustrate the effects of a change in the
dust mixture, I have run a series of calculations using Draine's
model corresponding to $R_V$=5.5. As shown in Fig.~\ref{fig:draine},
the three dust parameters increase at all wavelengths.  In the single
scattering approximation, the consequences on $B$ and $V$ LE
luminosities and colours can be anticipated from the following
considerations. Neglecting the small variation in the forward
scattering degree (see below), the LE luminosity is expected to change
proportionally to $\omega\; C_{ext}$ (see Eq.~\ref{eq:f}). Using the
suitable values (see Table~\ref{tab:templates}) this turns into an
increase of 0.29 and 0.44 mags in $B$ and $V$ respectively, while the
$(B-V)$ colour becomes 0.15 mag redder.

When multiple scattering becomes important, the exact outcome is not
so obvious. As I have shown in Sec.~\ref{sec:mcres}, a dust optical
depth enhancement produces an increase in the LE luminosity until this
reaches some maximum value, after which absorption prevails on
scattering and the luminosity starts to decrease again. Therefore,
increasing the scattering cross section causes an earlier luminosity
saturation for any given dust density.

On the other hand, an albedo increase translates into smaller losses
due to absorption, and therefore into a brighter light
reverberation. The same mechanism tends also to delay the luminosity
saturation, since higher values of the albedo turn into a smaller
efficiency of multiple scattering in absorbing the radiation. The
effect is going to be more pronounced at higher densities since, in
those conditions, a fair fraction of the photon packets which reach
the observer have undergone multiple scattering and therefore their
weight, given by Eq.~\ref{eq:weight}, strongly depends on the albedo.
Finally, an higher forward scattering degree tends to increase the
luminosity, especially at early phases, and to enhance the time
dependency (see Sec.~\ref{sec:analyt}).

\begin{figure}
\centering
\resizebox{\hsize}{!}{\includegraphics{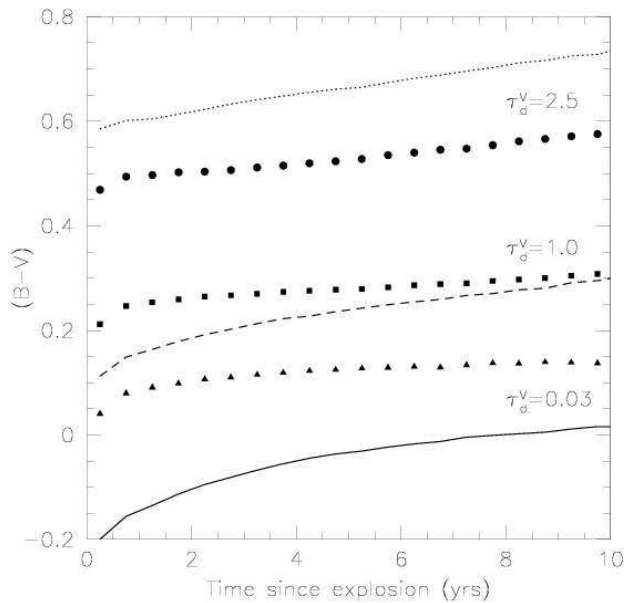}}
\caption{\label{fig:bmvRV55} Multiple scattering $(B-V)$ colour curves for the
thin shell case and for $R_V$=5.5 (symbols). The densities have been
reduced in order to give the same optical depths $\tau_d^V$ as in
Fig.~\ref{fig:bmv}. For comparison, the corresponding solutions for $R_V$=3.1
are also plotted (lines).}
\end{figure}

To illustrate the combined action of these mechanisms, in
Fig.~\ref{fig:bmvRV55} I have plotted the comparison between the
$(B-V)$ colours obtained for $R_V=3.1$ (lines) and $R_V=5.5$
(symbols), for the thin shell case ($R$=20 lyrs, $\Delta R$=2
lyrs). For the high $R_V$ solutions, the densities $n_0$ were adjusted
in order to produce the same optical depths $\tau_d^V$, i.e. 0.03, 1.0
and 2.5. As expected, for the low density case, the $R_V$=5.5 solution
(triangles) is about 0.15 mags redder than the $R_V$=3.3 one (solid
line), the difference being larger at earlier phases due to the
forward scattering effect. Then, at higher densities, the two
solutions (squares and dashed line) become similar, implying that the
optical depth effect is weakened by the albedo effect. Finally, for
$\tau_d^V$=2.5 (circles and dotted line) the $R_V$=5.5 solution is
bluer than the $R_V$=3.1 one. As a consequence, it is not possible to
deduce the optical depth {\it and} the extinction law from the observed LE
spectrum, even in the hypothesis the unreddened SN spectra are
available. At least one of the two has to be assumed in order to derive
the other. In fact, at least for the two different dust mixtures we
have used here ($R_V$=3.1 and 5.5), one can produce very similar
results just using different values for the LE optical depth.

\section{Light echo polarization}
\label{sec:polar}

Due to dust scattering, resolved LEs are expected to show strong
linear polarization, which should reach its maximum degree for the
dust which produces a scattering angle $\theta$=$\pi/2$
(\citealt{sparks94}, 1997). In single scattering approximation, this
occurs for dust lying on the SN plane, at a distance $ct$ from the SN
itself.

In this section, I want to address the effect of multiple scattering
on the observed polarization. A similar problem, even though in rather
different contexts, has been approached using MC techniques by several
authors (see for example \citealt{hoeflich}, \citealt{fischer}, 
\citealt{code}, \citealt{wood}, Bianchi et al. 1996), 
and here I basically follow the same procedure they have adopted.


The fundamental concepts are those exposed by Chandrasekhar (1950,
Chap.~I; hereafter C50). In this view, a photon packet is characterized by 
a Stokes vector $\bmath{S}=(I,Q,U,V)$, which at each scattering event undergoes
a transformation described by the following equation 
(C50, Eq.~210):

\begin{equation}
\label{eq:stokes}
\bmath{S} = \bmath{L}(\pi-i_2) \cdot \bmath{R} \cdot \bmath{L}(-i_1) \cdot \bmath{S_0}
\end{equation}

where $\bmath{S_0}$ denotes the incoming Stokes vector, $\bmath{R}$ is the
so-called scattering matrix and $\bmath{L}$ is a Mueller rotation matrix 
(see e.g. \citealt{tinbergen}, Chap.~4) which
transforms the Stokes vector from one reference system to the other.
Following C50, I use as reference plane that defined by the photon
direction and the $z$-axis, so that the required transformations are a
counter-clock rotation of amplitude $i_1$, to bring the Stokes
components into the scattering plane, and a clockwise rotation of
amplitude $\pi-i_2$ to recompute the Stokes components with the new
reference plane (see \citealt{vdhulst}). 

The explicit terms for $\bmath{L}$ can be derived from C50, Eq.~186 (see
for example \citealt{code}, Eqs.~3), while approximate 
expressions for $\bmath{R}$ terms in the case of H-G phase 
function can be found in \citet{white}.

\subsection{Single scattering polarization}

The effect of a single scattering on an unpolarized photon packet
($Q_0=U_0=V_0=0$) emitted by the central source, can be immediately
computed using the approximate expressions found by
\citet{white} and this gives

\begin{equation}
\label{eq:polsparks}
P_l= p_l\;\frac{1-\cos^2 \theta}{1+\cos ^2\theta}
\end{equation}

where $p_l$, the peak linear polarization, is of the order of 0.5. This result,
first applied to LEs by Sparks (1994, Eq.~10), implies
that the polarization is maximum for $\theta$=$\pi/2$, i.e. on the SN plane.

As for the polarization angle, it is easy to verify that the scattering 
plane and the electric field are perpendicular. As pointed out by 
\citet{sparks94}, this implies that the polarization direction is, at any given
position, tangent to the circle centered on the SN, and this can be
used as a distinctive feature to identify a LE.

Using Eq.~\ref{eq:costheta} in Eq.~\ref{eq:polsparks}, one can easily
compute the expected radial polarization profile (see
Fig.~\ref{fig:polprof}, solid line), which can be used to compare with
the multiple scattering solutions I am going to present in the next
sub-section.  I note that in single scattering approximation the
polarization degree is independent from $n_0$, $g$, $C_{ext}$ and
$\omega$.

\subsection{Depolarizing effect of multiple scattering}
\label{sec:mulpol}

The implementation of polarization calculations in a MC code is fairly
simple and it does not differ substantially from the one I have used
to compute the echo surface brightness.  After generating an
unpolarized photon packet, its polarization state is followed along
its random walk through Eqs.~\ref{eq:stokes}. In order to take into
account the forced nature of first scattering, the initial Stokes
vector is set to $\bmath{S_0}=(W_0,0,0,0)$, where $W_0=1-exp(-\tau_1)$
(see Sec.~\ref{sec:forced}).

As usual in MC methods (see \citealt{kattawar}), the
post-scattering Stokes parameters are weighted by the distribution
which they were sampled from, which in this case is $P_1$ itself,
i.e. the H-G phase function. In our case, all $P_i$
($i$=1,2,3,4) terms of the scattering matrix contain $P_1$ as
multiplicative factor (see \citealt{white}) and, therefore, this
corresponds to setting $P_1$=1.  Finally, in
order to account for the albedo effect and to make the radiation
transfer consistent with the one I have used for the luminosity
calculations, one would need to normalize the new Stokes parameters so
that

\begin{equation}
\label{eq:albedo}
I=\omega\; I_0
\end{equation}

as done, for example, by \citet{fischer}. Rigorously speaking, this
is true for the first scattering only, i.e. when the incoming light is
unpolarized. In fact, after the first scattering episode, the light
acquires some polarization degree and the fraction absorbed by the
next event does depend on the direction of the electric field with
respect to the scattering plane, through the following relation

\begin{equation}
\label{eq:polalbedo}
I = \omega\;[I_0 - p_l \frac{1-\mu^2}{1+\mu^2} \; (c_1Q_0 + s_1U_0) ]
\end{equation}

which is, in general, different from Eq.~\ref{eq:albedo}. Depending on
the scattering circumstances, this implies that the effective albedo
can be higher or lower than $\omega$, making this inconsistent with
Eq.~\ref{eq:albedo}, which I have {\it de facto} used in the
luminosity calculations. However, the simulations show that this
inconsistency applies to photon packets individually, while the global
effect tends to cancel and the average intensity ratio $I/I_0$
coincides with $\omega$, which can be therefore regarded as the
average dust albedo (or the albedo for unpolarized light).  In my
calculations I have used Eq.~\ref{eq:polalbedo}, even if the results
are practically identical to those one obtains from
Eq.~\ref{eq:albedo}.

The normalized Stokes components after each scattering are used as input for 
the next one, until the packet escapes the dust system at a given projected 
position on the sky plane.

Since each photon packet is completely independent from the others (i.e.
there is no phase relation between them), the resulting Stokes parameters
are additive (CH50). This means that the polarization state at any given 
projected position can be obtained simply adding the Stokes vectors 
of all photon packets escaping from that position along a given direction.

\begin{figure}
\centering
\resizebox{\hsize}{!}{\includegraphics{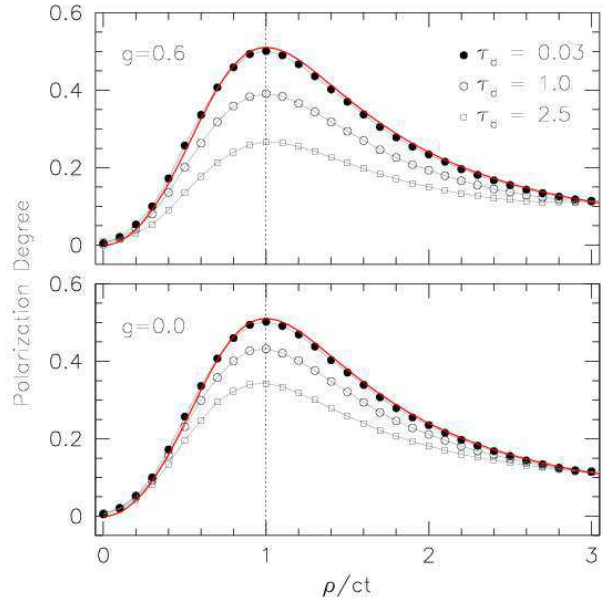}}
\caption{\label{fig:polprof} Radial polarization profiles for a homogeneous 
sphere ($R$=50 lyr) with forward ($g$=0.6, upper) and isotropic
($g$=0.0, lower) scattering, obtained for $t$=10 yrs.}
\end{figure}

In general, multiple scattering tends to act
as a depolarizing mechanism (see for example \citealt{code} and
\citealt{leroy}). In fact, successive scatterings occur with very
different and stochastic angular parameters and it is very improbable
that the polarization gained in the first scattering is increased.
As a consequence, the maximum polarization ring is expected to become less and
less marked, making its detection more and more difficult.

The results for a homogeneous sphere with $R$=50 lyr, which is
intended to mimic the case of SN immersed in a homogeneous medium, are
presented in Fig.~\ref{fig:polprof} for different values for the
density in the case of isotropic and forward scattering ($g$=0.6). The
agreement with the analytical solution is very good for low optical depths. 
As anticipated, for increasing density values the
polarization degree tends to decrease and the effect tends to be more
pronounced for forward scattering.  

It must be noticed that in all cases where the dust shows an axial
symmetry around the line of sight, the net linear polarization one
would measure in an unresolved LE image is null. Conversely, a net
polarization degree significantly different from zero would signal the
presence of asymmetry in the dust distribution, similarly to the case
of an asymmetric SN envelope (see for example \citealt{hoeflich}).  To
illustrate this effect, I have calculated the net polarization one
would observe from an unresolved uniform sphere placed at a distance
$D$=$R$=50 lyr from the SN and seen with a viewing angle $\alpha$=0,
$\pi/4$ and $\pi/2$, which turn out to be $<$0.1\%, 0.8\% and 6.0\%
respectively, for the usual values of relevant parameters ($n_0$=1
cm$^{-3}$, $\tau_d$=0.03, $g$=0.6, $\omega$=0.6). The observed
polarization angle, defined as the direction of the electric field
vibration, will then give an indication about the direction of maximum
asymmetry, provided that linear polarization is significant with
respect to the typical measurement errors.

Finally, after the second scattering, when the photon packets have
acquired some degree of linear polarization, they may gain also some
circular polarization. However, the simulations show that it is always
$P_c<$5$\times$10$^{-4}$, which is within the noise of our
calculations. Similar results were found by \citet{code}. For this
reason circular polarization can be safely neglected (e.g. $V$=0).

\section{Discussion}
\label{sec:discuss}

As I have shown, a LE with an optical luminosity $\sim$10 mag fainter
than the SN at maximum can be produced by a Ia in the presence of dust
with rather normal densities (1--10 cm$^{-3}$) and within 50--100 lyr
(see Sec.~\ref{sec:analyt}). The resulting integrated light curves
have typical slopes of 0.1--0.3 mag yr$^{-1}$. This becomes much higher
if the dust is very close to the SN, as it is the case, for instance, of a
stellar wind.

For a fixed distance between the SN and a given scattering element,
the contribution to the total flux is not constant and it depends on
its $z$ coordinate (see Fig.~\ref{fig:geom}), due to the presence of
forward scattering, which makes the process more efficient for
foreground dust. Moreover, at any given time, the echo luminosity
depends on the portion of the dusty region intersected by the
iso-delay surface. For example, a spherical cloud placed in front of
the SN at a distance $D$ along the $z$ axis is much more efficient
than an identical cloud at the same distance $D$ but with its center
located on the SN plane.

Of course, the two different geometries would produce very dissimilar
echo resolved images, but in the lack of LE resolved images, it is
very difficult to decode the dust geometry from the integrated light
curve alone.  In fact, unless the dust is confined within a region
with a characteristic dimension $R<$ct, even very different geometries
produce quite similar light curves (see Sec.~\ref{sec:analyt}).

The conclusion is that from the observed integrated light curve it is
not possible to deduce the dust distribution, since the light curve
shape is almost geometry independent and the echo luminosity can be
tuned simply changing the dust density and its distance. Therefore,
one is left with the question whether there are lots of dust far from
the SN or a small amount of dust close to the SN.

One may argue that an increase in the dust optical depth would
also produce observable reddening in the SN itself. This is true only
under some particular conditions. For example, it could well be that
there is a distant and optically thick cloud in front of the SN
responsible for a significant reddening (but irrelevant for the light
echo) while the scattering into the line of sight is produced by
off-centered dust. A good example of such a case is given by
SN~1989B, which was highly reddened but for which there is no clear LE
detection (\citealt{milne}). As I will show in a forthcoming paper,
where I will apply the results presented here to the known cases,
SN~1998es had a similar behaviour.  Due to the way the LE forms
(Sec.~\ref{sec:single}), one can have in fact several possibilities
and the LE detection can be accompanied by SN extinction or not, in
the same way that a heavy SN reddening is not necessarily related to
the actual onset of a LE.

As a matter of fact, LE properties depend on
the global dust geometry, while the SN reddening is linked to the dust
lying on the line of sight only. This is why $\tau_d$ and $\tau_{eff}$
have, in general, two different meanings, which tend to coincide only
under special conditions.  Therefore, unless one has LE
resolved images and an estimate of the distance $d$ to the SN, the
integrated light curve alone is not enough to answer our original
question, i.e. whether the SN is close or far from a dusty region.

Fortunately, as we have seen, the inclusion of multiple scattering
shows that the LE colour traces pretty well the effective dust
optical depth (Sec.~\ref{sec:mcres}). So, if an observed colour
including a reddening sensitive passband (i.e. blue) is available, one
can draw some conclusions about the dust density and the importance of
multiple scattering.  In fact, for low dust optical depths
($\tau^B_{eff}\leq$0.3), the resulting $(B-V)$ colour is similar to
the intrinsic colour of the SN at maximum light and it is practically
independent from the geometry (Sec.~\ref{sec:attenuation}). As a
consequence, any redder observed colour would signal the presence of an
higher optical depth.  In turn, this would give an additional
constraint when one tries to reproduce the observed echo luminosity,
thus reducing the geometry/density degeneracy which, however, can be
eliminated only when high resolution imaging is also available. 

Under these favorable conditions one can hope to derive a 3D mapping
of the SN surroundings, given the univocal relation between the
distance $r$ of the scattering elements from the SN and their impact
parameter $\rho$ (see Eq.~\ref{eq:eqdist}).

It is true that, as MC simulations show, this relation tends to break
down as the dust optical depth grows, since the phenomenon tends to
acquire a non-local nature. The exact details depend on the dust
geometry, but in general this takes place at pretty high values of
$\tau_{eff}$.  In this respect, the onset of multiple scattering is
much more important for the resulting LE luminosity, which does not
increase monotonically as the dust density grows
(Sec.~\ref{sec:mcres}), a fact which is foreseen also in the single
SSA approximation (Sec.~\ref{sec:ssa}). Since the effect is more
pronounced at bluer wavelengths, this produces the colour effect I
have just discussed. Speaking about this, it is worth noting that it
is generally assumed that the LE has a colour similar to that of the
SN at maximum light (corrected for reddening if there is any). As I
have shown (Sec.~\ref{sec:colour}), this is surely true when multiple
scattering is negligible but, if the SN is surrounded by a dense dusty
medium, both the SN and the echo are expected to be reddened, possibly
by different amounts, since in general $\tau_d \neq \tau_{eff}$, so
that one can have both an unreddened SN coupled to a red echo or a
reddened SN associated with a blue echo.  Both SNe 1991T
(\citealt{schmidt}) and 1998bu (\citealt{capp}) are examples of the
latter instance. In fact, the integrated $(B-V)$ colour of the LE was
about $-$0.1, while both SNe certainly suffered from significant
extinction. The fact that the LEs are definitely blue seems to suggest
that the SN reddening is generated by a front cloud whose
characteristic dimension $R$ is smaller than $\sqrt{ct(2ct+D)}$ (where
$D$ is the distance of the cloud from the SN itself), so that it does
not contribute to the LE for $ct>-D+\sqrt{D^2+R^2}$, while the LE is
generated by a less optically thick cloud. I will come back to this
point in the forthcoming paper.

Besides the optical depth, which is directly related to $C_{ext}$ and
plays a relevant role, dust albedo $\omega$ and forward scattering
degree $g$ are also important. For this reason, the exact results for
LE spectra and broad band colours depend on the dust mixture one
adopts. As a first approximation, in my simulations I have used the
canonical $R_V=3.1$ Milky Way composition \cite{draine}. A possible
further development of this work is the study of different dust
composition effects, similarly to what has been done by
\cite{sugerman03}.

Of course, the final results are expected to depend also on the SN
light curve one adopts as input.  All the simulations I have
presented in this paper made use of a template Ia input light curve,
derived from two {\it normal} events (e.g. SNe 1992A and 1994D). As I
have shown (see Sec.~\ref{sec:analyt}), this can be very well
approximated by a flash with a duration $\Delta t_{SN}$, a parameter
which normally changes with wavelength (see also
Table~\ref{tab:templates}). One may wonder what happens if this
parameter becomes larger, as it is the case for slow declining SNe
like 1991T ($\Delta t_{SN}^V$=0.116 yrs). The answer is easy and it is
given by Eq.~\ref{eq:Fsimple}, which was derived under simplifying
assumptions: integrated LE luminosity is proportional to $\Delta
t_{SN}$, and therefore slow declining events (e.g. 1991T-like) are
expected to produce brighter echoes than fast decliners. The effect is
mild, though, and the total range is of the order of 0.5 mag.

So far I have assumed that the distance $d$ to the SN is known, since
the main purpose of our inquiry is to derive the dust geometry. But
the paradigm can be in principle reversed, in the sense that the time
evolution of a resolved LE can be used to infer the distance. This was
attempted, for example, by \citet{bond} and \citet{tylenda} for the
case of V838 Mon, which has shown an impressive LE.  This method, in
principle promising because it gives a direct geometric distance
measurement, has nevertheless a weak point. In fact, it requires the
geometry to be known in order to properly use the apparent echo
expansion to derive the distance (the discussion in
\citealt{tylenda}).  Different geometries can be of course
disentangled looking at the expansion evolution, but this requires
quite a long time baseline.

It must be noticed that this technique is applicable when the echo edges
can be clearly detected and measured at different epochs, as in the case 
of V838 Mon, whose distance (6-8 kpc) makes this feasible, while for 
a SN at a typical distance of 15 Mpc this becomes practically not a 
viable method.

Another application of LEs for direct distance determination is the
one suggested by \citet{sparks94}, which makes use of the strong
polarization ring generated by the dust placed at a distance $ct$ from
the SN on the $z$=0 plane (see Sec.~\ref{sec:polar}). Provided that
there is dust on the plane of the SN, this allows one to get distance
estimates directly from polarimetric imaging. Since the maximum
polarization ring obviously expands at the speed of light and not at
the superluminal speed expected for the LE edges, this method is
practically applicable to old SNe only ($t>$50 yrs), always provided
that HST polarimetric data is available.

There are two problems that could hinder this technique. The first is
that what counts for the polarization detection is the polarized flux
which, if any, is smaller than the total flux. As I have shown, for a
uniform dust distribution (which I have mimicked with a sphere
centered on the SN) the total flux decreases at a rate of several
tenths of a magnitude per year during the first years, just because of
the efficiency loss due to forward scattering. This adds up to
something of the order of several magnitudes in 50 years, making the
detection of the polarized signal very difficult, even if its
polarization degree remains constant. The effect is going to be
worsened by the probable presence of a bright galaxy background, which
would certainly introduce additional noise in the polarimetric
measurements.  Secondly, the presence of multiple scattering tends to
depolarize the scattered light, again decreasing the chances of
detection. Besides this, I should as well note that a poor imaging
resolution tends to reduce the polarization one would actually measure
(see for example
\citealt{leroy}).

As I have mentioned, the full applicability of this method relies on
the presence of dust on the SN plane. If this is not the case, for
example because the region around the central source is empty within a
radius larger than $ct$, one can obtain only a lower limit for the
distance. An example of such a situation is given by V838 Mon (\citealt{bond}).
In general, even though the method is certainly promising and worthwhile of
being pursued, I am persuaded that the use of LEs
to study the SN environment is more rewarding than distance determination.
After all, deriving the distance to the SN with a factor 2 precision
is nowadays not a great progress, while the same accuracy on the estimate
of the distance of a Ia from a molecular cloud would be of high interest,
especially if this turns out to be small.

I like to conclude this discussion with the following consideration. I
have mentioned several times that in pretty normal conditions, a Ia is
expected to produce a LE about 10 magnitudes fainter than its
luminosity at maximum.  This is actually what happened for the two
known cases of SNe 1991T (\citealt{schmidt}) and 1998bu
(\citealt{capp}) which, at around 2 years after the explosion, clearly
deviated from the usual radioactive decay.  As a matter of fact, as a
result of a systematic LE search including 64 historical SNe,
\citet{boffi} have reported 16 possible candidates, only one of which
is a genuine Ia, i.e. SN~1989B (but see also \citealt{milne}).
Therefore, one may first inquire why only two events have been
detected and immediately conclude that this is simply because in the
vast majority of the cases there is not enough dust around Ia's, not
an unexpected conclusion for supposedly long-lived and small mass
progenitors.
 
At least for resolved HST images, this point has been thoroughly
addressed by \citet{sugerman03} in his extensive work on the
observability of LEs, and his conclusion is that {\it many echoes were
lost within the surface brightness fluctuations of background
unresolved starlight}. In the case of ground based observations, I
must notice that there are only a few Ia's observed at more than one
year past maximum. In fact, this has always been a problem, both due
to their faintness and to the presence of the host galaxy background.
Therefore, it is difficult to give a final answer on the basis of the
current list of LE detections and we will probably have to wait a bit
more in order to have a statistically significant sample.  This is
even more true if only over-luminous SNe are associated with dusty
regions.

What I can exclude here, at least on the basis of my calculations, is
that very dense ($n_0>$30 cm$^{-3}$), very close ($r<$50 lyr) and very
extended ($R>$30--50 lyr) clouds are not present in SNe of type Ia,
for they would produce well observable LEs, which is seemingly not the
case. Also, from the results shown in Sec.~\ref{sec:numeric} for the
double exponential galactic disk, it seems that Type Ia SNe tend to
explode far from the host galaxy disk, at vertical distances from the
galactic plane that are significantly larger than the dust height
scale $Z_d$, otherwise they would produce, again, detectable LEs.

\section{Conclusions}
\label{sec:conclusions}

In this paper I have discussed the phenomenology of unresolved LEs,
with particular attention to the effects of multiple scattering on
integrated luminosity and broad band colours. Even if the treatment is
absolutely general and it can be applied to LEs produced by any
transient source, I have devoted my analysis to the SN Ia case, due to
its relevance in the possible relation between the observed SN
properties and the explosion environment.  The main results I have
obtained can be summarized as follows:

\begin{itemize}
\item A type Ia SN is expected to produce a LE $\sim$10 mag fainter than the
SN at maximum when it is immersed in a medium with dust densities of the
order of 1 cm$^{-3}$;

\item If there is dust on the line of sight, the LE takes place immediately
after the SN outburst, but it becomes visible only when the SN has faded
away. This typically happens at 2-3 yrs past explosion;

\item In those cases one should rather indicate the phenomenon as a light 
reverb to make the parallel with acoustic physics more pertinent;

\item In general, the dust optical depth as seen from the SN along the line
of sight, $\tau_d$, is not sufficient to describe the integrated LE 
properties. A better description is given by $\tau_{eff}$, which is a sort
of weighted LE optical depth;

\item The integrated LE light curve tends to decline with time, both due
to geometrical effects and forward scattering. Higher forward
scattering tends to produce faster declines;

\item For typical values of the forward scattering degree ($g$=0.6) the
decline rate during the first few years is of the order of 0.1$-$0.3
mag yrs$^{-1}$;

\item It is not possible, on the basis of integrated LE curves alone, to deduce
the dust geometry. This can be inferred only if LE resolved images are
available;

\item For low values of $\tau_{eff}$ ($\la$0.3) the LE phenomenon can be
safely described with single scattering approximation.;

\item For higher optical depths, auto-absorption becomes relevant and 
attenuation ($\tau_{eff}<$1) or, better, multiple scattering must be included;

\item Multiple scattering has a number of effects, but the most important is
the LE reddening, which does not strongly depend on the dust geometry;

\item At low dust optical depths ($\tau_{eff}\la$0.3), the LE $(B-V)$ colour 
is similar to the intrinsic colour of the SN at maximum. The
importance of multiple scattering can be judged on the basis of
observed LE colors, which can give approximate values for
$\tau_{eff}$;

\item The LE spectrum is expected to change very slowly with time. It does
not resemble any of the SN spectra, being a mixture of all SN spectra
between $t$=0 and the time of observation;

\item Being the product of scattering, a LE is expected to be highly
polarized, with maximum polarization ($\sim$50\%) at geometrical
radius $ct$.  Multiple scattering tends to act as a depolarizer,
making the maximum polarization ring less and less pronounced;

\item In general the polarized flux, which is what counts for the polarization
detection, decreases with time. This hinders the application of
polarimetric imaging for direct distance determination.

\end{itemize}

\section*{Acknowledgments}
I am profoundly indebted to D. Di Bessoi for extensive and instructive
discussions. I also wish to thank E. Cappellaro for reading the
manuscript and for his helpful suggestions and ideas and I am grateful
to V. Utrobin, for his introduction to multiple scattering MC
simulations and to S. Bianchi and R. Chevalier for their very kind
help.  Finally, I like to thank the referee, Dr. B. Sugerman, for
pointing out a number of weak points and for suggesting several ways
of improving the quality of this paper.


 \bsp \label{lastpage} \end{document}